\documentclass[conference,compsoc]{IEEEtran}

\usepackage{microtype}
\usepackage{graphicx}
\usepackage{subfigure}
\usepackage{booktabs} 
\usepackage{multirow}

\usepackage{hyperref}
\usepackage{caption}
\usepackage{xcolor}
\usepackage{color}

\usepackage{amsmath,amssymb,amscd,url,enumerate, mathtools}
\usepackage[utf8]{inputenc}
\usepackage{wrapfig}
\usepackage{pdflscape}
\usepackage{tikz}
\usetikzlibrary{decorations.markings,arrows}
\usetikzlibrary{positioning,arrows,fit}

\usepackage{tikz-cd}
\usepackage{amsfonts}
\usepackage{amsmath}
\usepackage{graphicx}
\usepackage{mathrsfs}
\usepackage{dsfont}
\usepackage{enumitem}
\usepackage{array}
\usepackage{verbatim}
\newcolumntype{?}{!{\vrule width 1pt}}
\usepackage[capitalize,noabbrev]{cleveref}
\usepackage[font=footnotesize,labelfont=bf, labelsep=period, textfont=sl]{caption}
\usepackage{pythonhighlight}

\newcommand{\para}[1]{{\vspace{2pt} \bf \noindent #1}}  
\newenvironment{packed_itemize}{
\begin{list}{\labelitemi}{\leftmargin=1em}
\setlength{\itemsep}{1pt}                                                           
\setlength{\parskip}{0pt}                                                                                 \setlength{\parsep}{0pt}                                                                                  \setlength{\headsep}{0pt}                                                                                 \setlength{\topskip}{0pt}                                                                                 \setlength{\topmargin}{0pt}                                                                               \setlength{\topsep}{0pt}                                                                                  \setlength{\partopsep}{0pt}                                                                               }{\end{list}}

\DeclarePairedDelimiter\floor{\lfloor}{\rfloor}

\usepackage[strict]{changepage}

\usepackage{framed}

\definecolor{formalshade}{rgb}{0.95,0.95,1}

\newcommand{\kyb}{{\texttt{KYBER}}}

\newcommand{\fancyb}{{\mathfrak{B}}}
\newcommand{\binsec}{{\fancyb^{+}}}
\newcommand{\ternsec}{{\fancyb^{-}}}
\newcommand{\binomsec}{{\fancyb^{\eta}}}
\newcommand{\normsec}{{\mathcal{N}({\sigma})}}

\newcommand{\binsech}{{\fancyb^{+}_h}}
\newcommand{\ternsech}{{\fancyb^{-}_h}}
\newcommand{\binomsech}{{\fancyb^{\eta}_h}}
\newcommand{\normsech}{{\mathcal{N}({\sigma})_h}}

\definecolor{lightgreen}{HTML}{117733}
\definecolor{lred}{HTML}{FF0000}

\newcommand{\ejw}[1]{\textcolor{black}{#1}}

\begin{document}
\title{Benchmarking Attacks on Learning with Errors}

\author{\IEEEauthorblockN{Emily Wenger\IEEEauthorrefmark{1}\IEEEauthorrefmark{2},
Eshika Saxena\IEEEauthorrefmark{1},
Mohamed Malhou\IEEEauthorrefmark{1}\IEEEauthorrefmark{3}, 
Ellie Thieu\IEEEauthorrefmark{4} and
Kristin Lauter\IEEEauthorrefmark{1}}
\IEEEauthorblockA{\IEEEauthorrefmark{1}Meta AI, \IEEEauthorrefmark{2}Duke University, \IEEEauthorrefmark{3}Sorbonne Université\textsuperscript{\textparagraph}
\IEEEauthorrefmark{4}University of Wisconsin-Madison}}


\maketitle
\begingroup\renewcommand\thefootnote{\textparagraph}
\footnotetext{CNRS, LIP6, F-75005 Paris, France }

% TODO REMOVE THESE BEFORE SUBMISSION!!!!!!
\thispagestyle{plain}
\pagestyle{plain}

\begin{abstract}
Lattice cryptography schemes based on the learning with errors (LWE) hardness assumption have been standardized by NIST for use as post-quantum cryptosystems, and by HomomorphicEncryption.org for performing encrypted computations on sensitive data. Thus, understanding their concrete security is critical. Most work on LWE security focuses on theoretical estimates of attack performance, which is important but may overlook attack nuances arising in real-world implementations.  The sole existing concrete benchmarking effort, the Darmstadt Lattice Challenge, does not include benchmarks relevant to the standardized LWE parameter choices\textemdash such as small secret and small error distributions, and Ring-LWE (RLWE) and Module-LWE (MLWE) variants. 
To improve our understanding of concrete LWE security, we provide the first benchmarks for LWE secret recovery on standardized parameters, for small and low-weight (sparse) secrets.  
We evaluate four LWE attacks in these settings to serve as a baseline: the Search-LWE attacks uSVP~\cite{ADPS}, SALSA~\cite{stevens2024fresca}, and Cool\&Cruel~\cite{nolte2024cool}, and the Decision-LWE attack:  Dual Hybrid Meet-in-the-Middle (MitM)~\cite{Cheon_hybrid_dual}.  
We extend the SALSA and Cool\&Cruel attacks in significant ways, and implement and scale up MitM attacks for the first time. For example, we recover hamming weight $9-11$ binomial secrets for KYBER ($\kappa=2$) parameters in $28-36$ hours with SALSA and Cool\&Cruel, while we find that MitM can solve Decision-LWE instances for hamming weights up to $4$ in under an hour for Kyber parameters, while uSVP attacks do not recover any secrets after running for more than $1100$ hours.  We also compare concrete performance against theoretical estimates.  Finally, we open source the code to enable future research. 
\end{abstract}


\section{Introduction}
\label{sec:intro}

A full-scale quantum computer would threaten the security of most modern public key cryptosystems. A quantum computer could easily solve the hard math problems\textemdash such as integer factorization\textemdash on which many of these systems are based. Consequently, the cryptographic community has sought to develop  {\em quantum-resistant} cryptosystems, based on hard problems that quantum computers cannot easily solve. 

Cryptosystems based on Learning with Errors (LWE) have emerged as leading contenders. The LWE problem is: given many instances of a random vector $a$ of dimension $n$ along with a noisy inner product of $a$ with a secret vector $s$  modulo $q$, recover $s$.  
The hardness of LWE depends on the dimension $n$, the modulus $q$, and the distributions from which the secret and the error are drawn. 

Several LWE-based cryptosystems, such as \texttt{CRYSTALS-KYBER}~\cite{crys_kyber} were standardized by NIST, the US National Institutes of Standards and Technology~\cite{nist2022finalists} in 2024 for industry use in post-quantum cryptography. Furthermore, all publicly available fully homomorphic encryption (HE) libraries rely on the hardness of LWE for their security. HE was standardized by an industry consortium in 2018~\cite{HES}, with recent updated analysis given in~\cite{bossuat2024security}. Since standardization, LWE-based cryptosystems have been incorporated into applications such as the encrypted messaging protocol, Signal~\cite{signal_kyber}, and LWE-based HE schemes have been 
used in production by Microsoft in a password-breach detection scheme~\cite{msr_passwords}, and in the finance industry for encrypted computations on sensitive data~\cite{hint_sight}.

Consequently, vetting the security of deployed LWE-based cryptosystems  is critical. Although LWE is believed to be secure against attacks by a quantum computer, additional analysis is needed to ensure it is secure against classical (non-quantum) attacks. Numerous attacks against LWE have been proposed in recent years, but the LWE parameters proposed by NIST and HomomorphicEncryption.org are set to ensure $128$-bits of security in theory against all known attacks~\cite{nist_kyber_draft_standard, HES, bossuat2024security}. Such conclusions of security are often reached through use of the LWE Estimator~\cite{LWEestimator}, an open-source tool that estimates the resources needed to successfully attack given LWE parameters based on theoretical analysis and heuristic assumptions~\cite{nist_kyber_estimate, HES, bossuat2024security}.

Although the Estimator is a powerful tool, best practices in security analysis call for a multi-pronged approach to ensuring systems are secure. Given that billions of people may come to rely on LWE-based cryptosystems, additional avenues of assessing LWE security should be pursued. One obvious avenue is measuring concrete attack performance, to ensure that theoretical estimates line up with experimental observations. 
However, standardized methods for evaluating LWE attack performance are scarce. The sole exisiting concrete benchmark for LWE security is the Darmstadt Lattice Challenge~\cite{darmstadt_lwe}. 
While important for understanding the hardness of the Shortest Vector Problem (SVP), the Darmstadt challenges do not include benchmarks relevant to the standardized LWE parameter choices\textemdash such as small secret and small error distributions, and Ring-LWE (RLWE) and Module-LWE (MLWE) variants.

\para{Our Proposal: LWE Attack Benchmarking Challenge.} Given the importance of bolstering theory with experiments, and the current dearth of practical benchmarks for LWE attacks, we propose {\em the first practical LWE benchmarking challenge}. This challenge 
achieves three key objectives.  First, it complements existing theoretical estimates of LWE attack performance~\cite{LWEestimator}, 
replacing heuristic estimates with concrete running times and memory usage for known attacks. 
Second, it provides an avenue for measuring new attacks against existing attacks. Finally, it encourages the community to implement and optimize existing attacks, accelerating our understanding of concrete LWE security.

\para{Our Contributions.} We provide concrete benchmarks for LWE attacks with parameter choices from two key real-world use cases of LWE: \texttt{CRYSTALS-KYBER} and Homomorphic Encryption. We implement and evaluate {\em four} concrete LWE attacks\textemdash uSVP, SALSA, Cool\&Cruel, and Dual Hybrid MiTM\textemdash on these parameter settings. We then demonstrate successful secret recovery attacks on settings mimicking standardized LWE parameters, albeit with sparse secrets
(see Table~\ref{tab:all_results}):
$n=256$, $\log q =12$, Module-LWE of ranks $2$ and $3$ with binomial secret and error distributions for \texttt{KYBER}; and $n=1024$, $\log q = 26, 29$ with ternary secrets and narrow Gaussian error for HE.  

We start with sparse, or low weight, secrets, to benchmark attacks which succeed on these. The challenge is to successfully recover higher Hamming weight secrets, \ejw{moving towards recovery of general secrets.}  Prior work has demonstrated secret recovery for non-standard parameters, such as \ejw{small dimension $n=100-200$ and large $\log q$, with the goal of increasing $n$ or decreasing $q$} to make the LWE problem harder. Our proposed benchmark instead fixes $n$ and $q$ as the choices standardized for \texttt{KYBER} and HE, and identifies Hamming weight (or secret sparsity) as the sliding hardness parameter, to be increased to reach general secrets. 

Table~\ref{tab:all_results} shows secret recovery of hamming weight $9-11$ binomial secrets for \texttt{KYBER} parameters in $28-36$ hours with SALSA and Cool\&Cruel, while we find that MitM can solve Decision-LWE instances for hamming weights up to $4$ in under an hour for \texttt{KYBER} parameters, while uSVP attacks do not recover secrets after running for more than $1100$ hours.  
In the process of implementing and evaluating attacks, we made many new technical contributions, 
including: 
\begin{packed_itemize}
\item a new distinguisher to recover general (e.g. binomial, Gaussian) secrets from ML models in the SALSA attack;
\item a new linear regression algorithm to recover general secrets in the Cool\&Cruel~\cite{nolte2024cool} attack;
\item application of the RLWE cliff rotation approach 
 of~\cite{nolte2024cool} to attacks on RLWE in HE settings;
 \item application of the RLWE cliff rotation approach 
 of~\cite{nolte2024cool} to the Module-LWE setting for \texttt{KYBER}, introducing a ``cliff-splitting'' approach;
\item a method to preprocess Ring and Module LWE samples for use in SALSA and Cool\&Cruel attacks;
\item corrections to the Dual Hybrid MiTM attack, such as overestimates of number of short vectors needed and bad metrics for identifying secret candidates;
\item additions to the LWE Estimator.
\end{packed_itemize}
We also highlight interesting ``lessons learned'' from our attack implementations and experiments, including the importance of using a cryptographically sound random number generator to generate the random vectors $a$. We show recovery of much higher Hamming weight secrets when using a bad RNG. Finally, we open source our code for the attacks and evaluation. 

\para{Paper organization.} \S\ref{sec:lwe} discusses the Learning with Errors problem, including real-world use cases and proposed attacks. \S\ref{sec:benchmarks} describes prior LWE attack benchmarking work and introduces our benchmark settings. \S\ref{sec:attacks} provides details on the attacks we evaluate. \S\ref{sec:benchmark} presents our benchmark results. \S\ref{sec:lessons} highlights interesting lessons learned from implementing and evaluating attacks. \S\ref{sec:codebase} describes our open source codebase, and \S\ref{sec:discussion} lays out possible future work.

\section{Background on Learning with Errors (LWE)}
\label{sec:lwe}

\begin{table}[t]
    \centering
    \small
        \label{tab:notation}
     \resizebox{0.49\textwidth}{!}{
    \begin{tabular}{c@{\hskip 6pt}l}
        \toprule
        \textbf{Symbol} & \textbf{Description} \\
        \midrule
        $q$ & The modulus of the LWE problem considered \\
        $\mathbf s$ & The unknown secret, used to construct $b=\mathbf a \cdot \mathbf s + e$ \\
        $\chi_s$ & Distribution from which secret $s$ is chosen.  \\ 
        $\chi_e$ & Distribution from which error $e$ is chosen. \\ 
        $n$ & Dimension of vectors $\mathbf a$ and $\mathbf s$ or degree of polynomial \\ % removed LWE/RLWE/MLWE to save space! 
        $R_q$ & 
        Quotient Ring $\mathbb{Z}_q[X]/(X^n+1)$\\
        $\mathcal M $ & Module $R_q^k$\\

        $\text{Skew-Circ}(\mathbf{a})$ & Skew-circulant of a vector $\bf a$\\
        \bottomrule
    \end{tabular}
    }
    \caption{Notation used in this paper.}
    \vspace{-0.3cm}
\end{table}
 
The  Search-LWE problem is: given samples $(\mathbf{ A, b})$, where  $\mathbf{A} \in \mathbb{Z}^{m \times n}_q$ is a matrix with random entries modulo $q$, and $\mathbf{b} = \mathbf{A} \cdot \mathbf{s} + \mathbf{e} \in \mathbb{Z}^m_q$ are noisy inner products with a secret vector $\mathbf{s} \in \mathbb{Z}_q^n$ with error vector $\mathbf{e}$, recover the secret vector $\mathbf{s}$. 
 The Decision-LWE version of the problem is simply to decide whether ($\bf A, b$) are LWE samples or generated uniformly at random.
% $\mathcal{U}(0, q)$
The secret $\mathbf{s}$ and error $\mathbf{e}$ are chosen from distributions $\chi_s$ and $\chi_e$, respectively, and the hardness of the problem depends on $n$, $q$, and these distributions. 
We denote a single row of $({\bf A, b^t})$ as $(\mathbf{a}, b) \in \mathbb{Z}^n_q \times \mathbb{Z}_q$.

\subsection{LWE Settings}

\para{Secret and error distributions.} The LWE secret  $\mathbf{s}$  and error $\mathbf{e}$ distributions affect the hardness of LWE. Often, $\mathbf{s}$ and $\mathbf{e}$ are chosen from narrow (small) distribution to improve computational efficiency~\cite{HES}, although prior work demonstrates that narrow $\mathbf{s}$ distributions may be less secure~\cite{CCLS, bai_galbraith}. This work considers secret and error distributions where entries of the vectors are chosen as follows:
\begin{packed_itemize}
\item {\em Binary, $\binsec$}: uniformly from $\{0,1\}$.
\item {\em Ternary, $\ternsec$}:  uniformly from $\{-1,0,1\}$. 
\item {\em Binomial, $\binomsec$}: from a centered binomial distribution 
% parameterized by $\eta$ 
by taking independent uniform samples $(a_1 \ldots a_{\eta}, b_1 \ldots b_{\eta}) \leftarrow \{0,1\}^{2\eta}$, then outputting $\Sigma_{i=1}^{\eta} (a_i - b_i)$~\cite{crys_kyber}.
\item {\em Discrete Gaussian, $\normsec$}: from a normal distribution with mean $0$ and standard deviation $\sigma$, rounded to the nearest integer. 
\item {\em General, $\mathcal{U}(0,q)$}: uniformly from $\mathbb{Z}_q$.
\item {\em Fixed $h$ (secrets only):} any of the secret distributions above, but with a fixed number of nonzero coordinates $h$.
Binary secrets with $h$ nonzero coordinates denoted $\binsech$. 
\end{packed_itemize}

\para{Variants of LWE.} 
Variants of LWE such as Ring Learning with Errors (RLWE) have been proposed for use in real-world LWE applications. The HE Standard~\cite{HES} is based on RLWE, where a sample is defined by ($a(x), b(x) = a(x)s(x) + e(x)$) with $a(x), b(x)$ polynomials in a $2$-power cyclotomic ring  $R_q = \mathbb{Z}_q[X]/(X^n+1)$, $n$ is a power of $2$. In these rings, polynomial multiplication corresponds to matrix multiplication via the
coefficient embedding $\text{Emb}: R_q \rightarrow \mathbb{Z}_q^n$, $a(x)\rightarrow\mathbf a = (a_0, a_1, \dots, a_{n-1})$.
For $a(x)$ a random polynomial, the coefficients of the corresponding $b(x)$ can be obtained by multiplying a skew-circulant matrix $\mathbf{A} = \text{Skew-Circ}(\mathbf a)$ corresponding to $\mathbf a=\text{Emb}(a(x))$ by $\text{Emb}(s(x))$ and adding $\text{Emb}(e(x))$. 

Building on RLWE, yet another LWE variant is Module Learning with Errors (MLWE), which works in a free $R_q$-Module $\mathcal M = R_q^k$ of rank $k$. An MLWE sample is a pair ($\mathbf{a}, b$) where $\mathbf a = (a_1(x), a_2(x), \dots, a_{k}(x)) \in \mathcal M$, and $b = \mathbf {a\cdot s} + e \in R_q$ for some secret vector of polynomials $\mathbf s 
= (s_1(x), s_2(x), \dots, s_{k}(x))
\in \mathcal M$, and error polynomial $e$ chosen according to the specified distribution.

\subsection{LWE in the real world} 
\label{subsec:lwe_real}
LWE-based cryptosystems are used in two important real-world contexts: as part of the NIST-standardized set of post-quantum public-key encryption algorithms~\cite{nist2022finalists, crys_kyber}, and in homomorphic encryption (HE) applications~\cite{HES, bossuat2024security}. 

\para{Kyber.} The 5-year NIST competition to select algorithms for use in post-quantum cryptography chose an LWE-based cryptosystem  \texttt{CRYSTALS-KYBER} as a Key-Encapsulation Mechanism (KEM)~\cite{nist2022finalists}. \kyb{} is an MLWE system with binomial secrets and error distributions, $\binomsec$ whose parameters are listed in Table~\ref{tab:kyber_params}.

\begin{table}[h]
    \centering
    \begin{tabular}{ccccccc}
    \toprule
        $n$ & $k$ & $q$ & $X_{\bf  s}$ & $X_{\bf e}$ & $ \eta$ & NIST Security Level \\ \midrule
        $256$ & $2$ & $3329$ & $\binomsec$ & $\binomsec$   & 3 & 1  \\ 
        $256$ & $3$ & $3329$ & $\binomsec$  & $\binomsec$ & 2  & 3 \\
        $256$ & $4$ & $3329$ & $\binomsec$  & $\binomsec$ & 2  & 5 \\ 
        \bottomrule
     \end{tabular}
    \caption{Proposed standard KEM parameters for the Kyber MLWE scheme. NIST security levels $1$, $3$, and $5$ correspond to the expected security of brute-forcing AES-$128$, $192$, and $256$, respectively~\cite{nist_kyber_draft_standard}. }
    \label{tab:kyber_params}
    \vspace{-0.2cm}
\end{table}

\para{Homomorphic Encryption.} 
Most publicly available HE libraries implement RLWE-based systems and often use small (binary, ternary, narrow Gaussian), sparse secrets and small error. 
Small, sparse secrets enable bootstrapping for evaluating deep circuits required for deep neural nets, such as in encrypted AI model inference~\cite{gilad2016cryptonets}. 
In some proposed schemes, secrets have only $h=64$ active bits for dimension $n=2^{14}$~\cite{heaan}. 
Table 4.2 of~\cite{bossuat2024security} gives the highest modulus $q$ that can be safely used for a given $n$ with  ternary or narrow Gaussian secret distribution $\chi_{s} = \mathcal{N}( 3.19)$ and narrow Gaussian error $\chi_{e} = \mathcal{N}(3.19)$ (see Table~\ref{tab:fhe_params}).
\begin{table}[h]
    \centering
    \begin{tabular}{ccc}
    \toprule
        & \multicolumn{2}{c}{$\log_2 q$} \\ \cmidrule{2-3}
        $n$ & $\chi_{\bf s}$ = $\ternsec$ & $\chi_{\bf s}$ = $\mathcal{N}(3.19) $ \\ \midrule
        $1024$ &  $26$ & $29$ \\ 
        $2048$ & $54$ & $56$ \\ \bottomrule
    \end{tabular}
        \caption{Proposed standard parameters for RLWE-based Homomorphic Encryption schemes with security level $\lambda=128$.
         Error $\chi_e =\mathcal{N}(3.19)$
        ~\cite{bossuat2024security}.}  
    \label{tab:fhe_params}
    \vspace{-0.3cm}
\end{table}

\vspace{-0.2cm}
\subsection{Attacks on LWE} 
\label{subsec:known_attacks}
\vspace{-0.2cm}

\para{Lattice reduction.} Nearly all attacks on LWE rely in some way on {\em lattice reduction} algorithms, which  systematically project vectors onto linear subspaces to reduce their size and to find short vectors.
LLL~\cite{LLL}, is a polynomial time algorithm which produces short, nearly-orthogonal lattice bases. LLL is efficient but finds only exponentially bad approximations to the shortest vector. 
The Block Korkine-Zolotarev (BKZ) algorithm~\cite{BKZ} produces shorter vectors than LLL but runs in exponential time. BKZ variants like BKZ2.0~\cite{CN11_BKZ} and Progressive BKZ~\cite{xia2022improved, aono2016improved, wang2022improved} improve efficiency.  A subroutine of BKZ requires finding the shortest vector in a sub-lattice of smaller dimension.  One option for this subroutine is {\em sieving}, a technique which produces exponentially many short vectors in a small amount of time.  Although efficient sieving algorithms have been proposed and implemented~\cite{ducas_sieving, albrecht2019general}, sieving requires exponential memory. Recent work~\cite{ryan2023fast} proposed \texttt{flatter}, a fast lattice reduction algorithm that produces vectors with quality on par with LLL, using careful precision management techniques.

\para{LWE attacks.}  We consider 3 attacks on Search-LWE:  uSVP~\cite{HES}, machine learning (ML)~\cite{stevens2024fresca}, ``Cool\&Cruel"~\cite{nolte2024cool}, plus Decision-LWE
dual attacks~\cite{albrecht2017lattice,Cheon_hybrid_dual}. 

\vspace{0.05cm}
\noindent {\bf {\em uSVP attack.}} 
The uSVP attack constructs a lattice from the LWE samples $(\mathbf{A},\mathbf{b})$ in such a way that the secret vector $s$ can be recovered from the unique shortest vector in that lattice.  The attack relies on lattice reduction to find the shortest vector, and it only succeeds if the correct shortest vector is recovered.~\cite[Section 2.1.2]{HES}

%%% PUT HERE FOR SPACING.
% \usepackage{multirow}
\begin{table*}[t]

\begin{tabular}{cccccccc} 
\toprule
\multirow{2}{*}{\bf Paper} & \multirow{2}{*}{\bf Attack type} & \multicolumn{5}{c}{\bf Parameters} & \multirow{2}{*}{\bf \begin{tabular}[c]{@{}c@{}}Reported \\ performance \\metrics \end{tabular}} \\[0.1cm] \cmidrule{3-7}

 &  & \textbf{Setting} & $n$ & $\log_2 q$ & Secret distribution & Error distribution &  \\ 
 \midrule

\cite{CCLS} & Primal uSVP & LWE & $40 \le n \le 200$ & $7 \le \log_2 q \le 21$ & $\binsec$, $\ternsec$, $\normsec$, & $\mathcal{N}(0,3)$ & Time \\

\cite{xia2022improved} & Primal uSVP & LWE~\cite{darmstadt_lwe} & $40 \le n \le 90$ & $11, 12, 13$ & $U_q$ & $\mathcal{N}(0, 40 \le \sigma \le 64)$  & Time, Mem \\

\cite{postlethwaite2021success} & Primal uSVP & LWE & $72 \le n \le 100$ & $7, 9$ & $\binsec$, $\ternsec$ & $\binsec$, $\ternsec$ & Success \\

\cite{wang2022improved} & Primal uSVP & LWE~\cite{darmstadt_lwe} & 60, 75 & 12, 13 & $U_q$ & $\mathcal{N}(0, \sigma = 28, 36)$   & Time \\

\cite{dachman2020lwe} & BDD/uSVP & LWE & $n=70, 80$ & $12$ & $\mathcal{N}(0, 20)$ & $\mathcal{N}(0, 20)$ & Success \\ 

\cite{albrecht2019general} & uSVP & LWE~\cite{darmstadt_lwe} & $40 \le n \le 75$ & $11 \le \log_2 q \le 13$ & $U_q$  & $\mathcal{N}(0, 28 \le \sigma \le 48)$  & Time \\

 \cite{rumpsession} & MiTM & LWE & $256$ & $12$ & $\binsech$ & $\mathcal{N}(0, 3)$ & Time \\ 

\cite{budroni2020attacks} & MiTM & I-RLWE & $105 \le n \le 130$ & 21, 22 & $\mathcal{N}(0, \sqrt{n})$ & $\mathcal{N}(0, \sqrt{n})$ & Success \\
\cite{salsa} & ML & LWE & 128 & 9 & $\binsech$ & $\mathcal{N}(0, 3)$ & Time \\
\cite{picante} & ML & LWE & 350 & 32 & $\binsech$ & $\mathcal{N}(0,3)$ & Time \\ 
\cite{verde} & ML & LWE & 512 & 63 & $\binsech$, $\ternsech$ & $\mathcal{N}(0,3)$ & Time \\

\cite{stevens2024fresca} & ML & LWE & 512, 768, 1024 & 41, 35, 50 & $\binsech$, $\ternsech$ & $\mathcal{N}(0, 3)$ & Time \\ 

\cite{nolte2024cool} & Cool\&Cruel & LWE/RLWE & $256 \le n \le 768$ & $12$, $35$, $50$ & $\binsech$ & $\mathcal{N}(0, 3)$ & Time \\ \bottomrule

\end{tabular}
\caption{\textbf{Summary of all concrete evaluation results for attacks on Search and Decision LWE found in literature in last decade.} When Setting is ``LWE~\cite{darmstadt_lwe}", the attack was evaluated specifically on lattices from the Darmstdat challenge. ``Reported performance metrics'' refers to the attack performance metrics in the paper, as different evaluations report different metrics: Time = time to attack success, Mem = memory used in attack, and Success = whether an attack succeeded or not (used for papers where neither time nor memory are reported, but experiment results are included). }
\label{tab:concrete_overview}
\vspace{-0.2cm}
\end{table*}

\vspace{0.05cm}
\noindent {\bf {\em Dual attacks.}} Dual attacks~\cite{MR09} solve {\it the Decision-LWE problem} by finding a short enough vector in the dual lattice  $\Lambda = \{ x \in \mathbb{Z}^n_q | Ax = 0 \mod q \}$ using lattice reduction and/or lattice sieving~\cite{ducas_sieving, matzov}. Several variants of the dual attack work especially well for sparse binary secrets so we focus on those here: the Dual Hybrid and the Dual Hybrid Meet-in-the-Middle (MitM). 

The Dual Hybrid attack~\cite{Albrecht2017_sparse_binary} 
splits the matrix $\bf A$ into two parts and runs lattice reduction on one part and a guessing routine on the other. The basic version ``guesses'' that the columns in the second part of $\bf A$ correspond to all zero bits of the secret, so they don't contribute. This version is only relevant for sparse secrets.
The success probability of this method is very low, and the formula in~\cite[p.17]{Albrecht2017_sparse_binary} underestimates the number of times this attack would need to be run in order to succeed with high likelihood (the formula is correct but the approximation given is not).  To improve the success rate, one can either exhaustively guess all possible secrets in the second part, or use a MitM approach~\cite{Cheon_hybrid_dual}.  The exhaustive guess approach takes exponential time, while the MitM requires exponential memory.  The MiTM approach builds a table of partial secret guesses and queries it with other guesses to find a candidate for part of the secret.

\vspace{0.05cm}
\noindent {\bf {\em Machine learning (ML) attacks.}} The SALSA papers~\cite{salsa, picante, verde, stevens2024fresca} solve Search-LWE by training an ML model to predict $b$ given $\bf a$ for a fixed secret, and then use the model as an oracle to recover the secret key.
The SALSA attack first preprocesses a large amount of data (roughly 2 million samples, generated from $4n$ samples through repeated partial lattice reduction of random subsets).  
Encoder-only transformer models are trained on these datasets of reduced LWE samples ($\bf A$, $\bf b$).
As soon as the model learns to predict $\bf b$ from $\bf A$ with some accuracy, the secret can be recovered using special queries to the model. The attack recovers sparse binary and ternary secrets in dimension $n \le 1024$~\cite{stevens2024fresca}. 

\vspace{0.05cm}
\noindent {\bf {\em Cool\&Cruel (CC) attack.}} Recent work~\cite{nolte2024cool} leverages an experimentally observed ``cliff'' in reduced LWE matrices to recover sparse secrets. Like the ML attack, this attack first reduces a large set of LWE samples. It observes that the first columns of the $\bf A$ matrix remain unreduced (coordinates $\sim \mathbb{U}(0, q)$) after lattice reduction, while the remaining columns are reduced and have small norms.  The unreduced bits are ``cruel'' and reduced ones are ``cool''. The cool columns of $\bf A$ can be initially ignored in guessing secrets, and for some settings this reduces the search space far enough to make brute force feasible. After cruel secret bits are recovered, an efficient greedy algorithm is used to recover the ``cool'' bits.

\subsection{Prior Concrete Evaluations of LWE Attacks}

Many other LWE attacks and variants have been proposed, so one would expect that experimental evaluations of these attacks would be common. Unfortunately, this is not the case. Table~\ref{tab:concrete_overview} lists papers (last $\sim10$ years) giving concrete experimental results (e.g. time, memory requirements) for attacks on Decision or Search LWE. This paper list may not be exhaustive, but is complete to the best of our knowledge. Table~\ref{tab:github_repos} in the Appendix lists open-source Github repositories with implementations of attacks on Decision or Search LWE. This list is a subset of the first, as not all  papers with experimental results open-source their code.

The Darmstadt LWE Challenges, referred to in some of the entries in Table~\ref{tab:concrete_overview}, aim to benchmark LWE attack performance. However, the parameters are not relevant to LWE in practice since the  Darmstadt challenge are in small dimension $n \le 120$, and all secrets are chosen from the uniform distribution mod $q$.  The website recently announced that there was a bug in generation, resulting in challenges which were unsolvable. These challenges primarily explore SVP hardness as error and modulus change. 

Several things stand out from the lists in Table~\ref{tab:concrete_overview} and ~\ref{tab:github_repos}. First, they represent but a fraction of the many papers published on LWE each year. Most papers on LWE attacks provide theoretical estimates of attack performance rather than concrete evaluations. This is understandable, as many aim to understand the cost of attacking real-world LWE settings, and attacks for these settings should be computationally infeasible. Second, there is no discernible trend in the settings for concrete evaluation. A few use Darmstadt LWE challenges, but others choose LWE settings ad-hoc. Finally, few papers systematically compare different attacks' experimental performance, even for small parameter settings.

\vspace{-0.1cm}
\section{LWE Attack Benchmarks} 
\label{sec:benchmarks} 
\vspace{-0.1cm}

\begin{table*}[t]
\begin{minipage}{0.48\textwidth}
\centering
\begin{tabular}{ccclccc}
\toprule
$n$ & $k$ & $\log_2 q$ & $q$ & $X_s$ & $X_e$ & $\eta$ \\ \midrule %& $h$ \\ \midrule
256 & 2 & 12 & 3329 & $\binomsech $ & $\binomsec$ &  2  \\[0.2cm] %\midrule & $[6,10]$ 
256 & 2 & 28 & 179067461 & $\binomsech $ & $\binomsec$  &  2  \\[0.2cm] %\midrule & $[13, 19]$ 
256 & 3 & 35 & 34088624597 & $\binomsech $ & $\binomsec$  &  2  \\ \bottomrule % & $[11,17]$
\end{tabular}
\caption{Proposed Benchmark settings for Kyber. All use Module-LWE. $\eta=2$ matches  $\eta_1$ for Kyber-768 standard setting.}
\label{tab:kyber_bench}
%\vspace{-0.1cm}
\end{minipage}\hfill
\begin{minipage}{0.48\textwidth}
\centering
\begin{tabular}{cclccc}
\toprule
$n$ & $\log_2 q$ & $q$ & $X_s$ & $X_e$  \\ \midrule % & $h$
1024 & 26 & 41223389 & $\ternsech$ & $\normsec$  \\[0.2cm]
1024 & 29 & 274887787 &  $\ternsech$ & $\normsec$  \\[0.2cm]\
1024 & 50 & 607817174438671 & $\ternsech$ & $\normsec$  \\ \bottomrule
\end{tabular}
\caption{Proposed benchmark settings for HE. $\sigma=3.19$ for both $X_e$ and $X_s = \normsech$ (where appropriate). All settings use Ring-LWE.}
\label{tab:he_bench}
\end{minipage}
\end{table*}

\subsection{Benchmark Settings}

We propose a set of practical benchmark settings to encourage concrete evaluations of LWE attack performance, using the following criteria.
First, parameters should come from {\bf real-world choices}, such as narrow secret and error distributions and LWE variants like Ring- and Module-LWE, that are standardized or used in deployed LWE cryptosystems. Prior work shows that such parameter choices\textemdash including binary secrets, small error, and use of the ring-LWE variant\textemdash may weaken LWE, necessitating further experimental study~\cite{bai_galbraith, ELOS, CCLS, postlethwaite2021success, nolte2024cool}. The challenges should also have {\bf tunable hardness settings}. The hardness of LWE problems depends on dimension $n$, modulus size $q$, error distribution $\chi_e$, and secret distribution $\chi_s$. Generally, larger dimension $n$, smaller modulus $q$, and secret/error distributions with higher standard deviation ($\sigma_e$, $\sigma_s$) make LWE more difficult. A good set of challenges should fix some parameters and allow others to vary so that concrete attacks can be bench-marked to provide interesting insights.

Using these criteria, we propose two sets of benchmark settings for measuring concrete LWE attack performance based on two important real-world applications of LWE: \kyb{} and Homomorphic Encryption. Security guidelines have been proposed for each of these, outlining the required LWE distribution, $n$, $q$, $\chi_s$, and $\chi_e$ for real-world implementations, as discussed in \S\ref{subsec:lwe_real} and Tables~\ref{tab:kyber_params} and~\ref{tab:fhe_params}.  Since $n$, $q$, $\chi_s$, and $\chi_e$ are fixed for \kyb{} and for $n=1024$, $\chi_s$ and $\chi_e$ are fixed for HE, our challenges vary problem difficulty by changing the number of nonzero secret coordinates, $h$, and modulus, $q$. We divide our proposed benchmark settings into parameters focused on \kyb{}, using MLWE, and parameters focused on HE, using RLWE, and present the proposed settings in Tables~\ref{tab:kyber_bench}  and~\ref{tab:he_bench}. 

{\it The challenge is to solve LWE problems for larger $h$ in these standard parameter settings.} \ejw{Prior work has extensively explored attack performance for small $n$ with uniform secrets (e.g. ~\cite{darmstadt_lwe}). 
In addition, it is already well-known that lattice problems such as LWE are not hard when $\log q$ becomes large enough, for any fixed $n$~\cite{laine2015key,CCLS}.
Our challenge instead encourages implementation of full-scale attacks on large $n$ and small $q$, for small secret distributions which are standardized but potentially more risky.
We tune hardness by increasing the number of non-zero elements in the secret.}

\subsection{Choosing Attacks to Evaluate}

To choose which attacks to evaluate, we focused on: (1) attacks which have open-source implementations, (2) attacks which have already been evaluated in nontrivial settings (e.g. dimension $n > 256$),  (3) attacks that require less than $750$ GB of RAM (our machine capacity). 
We chose to evaluate the 4 attacks discussed in Section~\ref{sec:attacks}: uSVP, ML, and Cool\&Cruel for Search-LWE and Dual Hybrid MitM for Decision-LWE. 
Our attack evaluation provides a first set of benchmarks for the proposed settings. Future work should improve these and evaluate additional attacks. 

For the uSVP approach, we did not consider the DBDD attack~\cite{dachman2020lwe}, since it uses ``hints'' that other attacks do not have. This leaves us with the uSVP attack using Kannan's embedding (from the \cite{verde} codebase) as the only other open source option. We considered whether to use sieving or enumeration as the BKZ SVP-subroutine. Although fast sieving implementations are available, sieving memory requirements are exponential in $n$. For example, the GPU G6K lattice sieving implementation of~\cite{ducas_sieving} requires petabytes of data for $n > 160$ (see Appendix Table~\ref{tab:sieving_mem}). We lack such resources, so we use \texttt{fplll}'s BKZ2.0 with an enumeration SVP oracle.

For the ML attack, we use the open-source codebase from~\cite{verde}. We leverage the lattice reduction part of this codebase to process data for the cliff attack of~\cite{nolte2024cool}, since the pre-processing steps in these two attacks are the same. We base our cliff attack brute force secret guessing and greedy recovery algorithms on open source code from~\cite{nolte2024cool}. Finally, for Decision-LWE, we build on and scale up the dual hybrid MiTM attack implementation from~\cite{Cheon_hybrid_dual}.

\subsection{Evaluation Metrics} 
\label{subsec:metrics}
\vspace{-0.1cm}

 In Table~\ref{tab:all_results} we present the {\bf \em {attack time in hours}} corresponding to the {\bf {\em highest Hamming weight $h$}} secret recovered by each attack, along with the {\bf {\em compute resources used}} to conduct the attack.  Memory requirements are also important, and a limitation for scaling MitM, but for our comparisons in Table~\ref{tab:all_results} we run all attacks on the same machines with up to $750GB$ of RAM, and present memory requirements for MitM in Table~\ref{tab:both_mitm_mem}. 
 
 Some of the attacks can be parallelized, so for parallelizable parts of the attack, we report ``time $\cdot$ \#\{devices\}'', where \{device\} is CPU, GPU. For non-paralellizable parts, we report ``Single {device} time''. We then report the ``Total time (assuming full parallelization)''.  For each setting and attack, we experiment with different Hamming weight $h$ secrets, $10$ experiments per $h$. 

\para{Hardware specifics.} All attacks are run on 2.1GHz Intel Xeon Gold CPUs and/or NVIDIA V100 GPUs. Our machines have 750 GB of RAM, while the GPUs have 32 GB. All attacks we run must work within these memory limits. 

\section{Attack Implementations and Innovations}
\label{sec:attacks}

Here, we present details of the attacks we evaluate, as well as the innovations we introduce to make these attacks run on the proposed benchmark settings. Refer to the original attack papers for details.  All attacks are implemented in a codebase available at  \url{https://github.com/facebookresearch/LWE-benchmarking}.
Implementations are written in Python and leverage \texttt{fplll} and \texttt{flatter}~\cite{ryan2023fast} libraries for lattice reduction (with enumeration as the SVP oracle in BKZ2.0). Since all attacks rely on lattice reduction, benchmarking them using the same lattice reduction implementation allows for fair comparison.  Any improvement to lattice reduction would benefit all attacks. 

\subsection{uSVP}

We solve uSVP using Kannan's embedding~\cite{kaanan_extension}, as implemented in the~\cite{verde} open source codebase. The attack setup is as follows. Given an LWE sample $(\mathbf{ A, b}) \in \mathbb{Z}_q^{(m \times n)} \times \mathbb{Z}_q^m$, Kannan's embedding is constructed using the q-ary format suggested by~\cite{verde} to speed up reduction:
\[
\begin{bmatrix}
0 & qI_m & 0 \\ 
I_n & A^T & 0\\
%$0 & qI_m & 0 \\
0 & b & 1
\end{bmatrix}
\]

The space spanned by these rows contains an unusually short vector $\begin{pmatrix} s & -e & -1\end{pmatrix}$.

\noindent Thus lattice reduction recovers the secret once it finds the shortest vector in the lattice. 

\para{Implementation details.} Our implementation of uSVP multiplies the $I_n$ matrix by a factor $\omega$, which balances $s$ and $-e$ in the discovered short vector. $\omega$ is determined by formulae given in~\cite{CCLS}. We run lattice reduction using BKZ2.0~\cite{CN11_BKZ} and incorporate the improvements suggested in~\cite[Appendix A.7, p. 17]{verde}. Future improvements to our implementation of the uSVP attack could incorporate more advanced BKZ schemes, such as Pump and Jump~\cite{wang2022improved}. 

\subsection{ML Attack} 
\label{subsec:ml}

Our implementation of the ML attack is based on the open-source code of~\cite{verde}, incorporating the improvements from~\cite{stevens2024fresca}. The attack starts with $4n$ eavesdropped LWE samples $(\mathbf{A,b}) \in \mathbb{Z}_q^{4n \times n}, \mathbb{Z}_q^{4n}$. Then, a subsampling trick is employed to create many new LWE samples: select $m$ random indices from the $4n$ set to form  $(\mathbf{A}_i,\mathbf{b}_i) \in \mathbb{Z}_q^{m \times n}, \mathbb{Z}_q^m$. To ``preprocess'' this data and create a model training dataset, an important step of the ML attack, a q-ary embedding $\mathbf{\Lambda}_i$ of $\mathbf{A}_i$ is constructed via: 
\begin{equation}
\mathbf{\Lambda}_i =
 \begin{bmatrix}
0 & q\cdot \mathbf{I}_n \\
\omega\cdot\mathbf{I}_m & \mathbf{A}_i 
\end{bmatrix}
\label{eq:qary}
\end{equation}
 Lattice reduction on $\mathbf{\Lambda}_i$ finds a unimodular transformation $\begin{bmatrix} \mathbf{L} & \mathbf{R} \end{bmatrix}$ which minimizes the norms of $\begin{bmatrix} \mathbf{L} & \mathbf{R} \end{bmatrix} \mathbf{\Lambda_i}
= \begin{bmatrix} \omega \cdot\mathbf{R} & \mathbf{R}\mathbf{A}+q\cdot\mathbf{L} \end{bmatrix}$. $\omega$ is a scaling parameter that trades-off reduction strength and the error introduced by reduction. This $\mathbf{R}$ matrix is then applied to the original  $(\mathbf{A}_i,\mathbf{b}_i)$ to produce reduced samples $(\mathbf{RA}_i,\mathbf{Rb}_i)$ with smaller norms. Repeating this process many times (paralellized across many CPUs) produces a dataset of reduced LWE samples. 

This dataset is used to train a machine learning model $f$ to predict $\mathbf{R}b$ from input $\mathbf{Ra}$. If $f$ ever learns this task (even poorly), it has implicitly learned the LWE secret $\mathbf{s}$. At this point, a distinguishing algorithm is run periodically to extract $\mathbf{s}$ from $f$. This algorithm feeds special inputs to $f$ and discerns secret bit values from the model's response. 

\para{Implementation details.} The ML attack, as presented in~\cite{stevens2024fresca}, trains on LWE data and can recover binary and ternary secrets. Our improvements are: 1) adapting the attack to tackle the benchmarks proposed in this work, 2) introducing methods to reduce RLWE/MLWE samples as LWE samples, 3) exploiting the ``rotation'' trick on RLWE/MLWE data (first proposed in~\cite{nolte2024cool}), and 4) introducing a {\it slope distinguisher} to recover more general secrets (e.g. binomial and Gaussian).

\vspace{0.1cm}
\noindent {\bf {\em Reducing R/MLWE samples.}} 
Assume we start data preprocessing with $4n$ Module-LWE (RLWE samples are Module-LWE with $k=1$) samples, rather than $4n$ LWE samples. Treating these polynomial vectors as $kn$-long vectors of concatenated coefficients, we can employ the same sub-sampling trick as before. We sub-select $m$ vectors from the $4n$ sets to create an ``LWE-like'' matrix that is then reduced. Then, individual reduced rows can be circulated to create reduced MLWE samples, creating $kn$ reduced MLWE samples for the cost of one. 

\vspace{0.1cm}
\noindent {\bf {\em Rotation trick.}} As observed in~\cite{nolte2024cool}, LWE samples reduced using the embedding of ~\cref{eq:qary} have ``reduced'' and ``unreduced'' parts. When we reduce RLWE or MLWE polynomials, this behavior persists. In the $2$-power cyclotomic RLWE case, if a reduced polynomial $a(x)$ from a sample $(a(x), b(x))$ 
has a reduced part and unreduced part $a = (\mathbf a_u, \mathbf a_r)$, then 
the $n-l^{th}$ row in the skew-circulant matrix created from $a$ will exhibit a cliff shifted by $l$ positions. This pattern is replicated in each component in Module-LWE, see \cref{appx:cliffshift}. For sparse secrets, this represents a huge weakness since we can shift the unreduced region $\mathbf a_u$ around so that it corresponds to the sparsest region of the secret $\mathbf s$. 
 In practice, we train models on all $n$ possible shifted datasets and terminate when one model recovers the secret.

\begin{figure}[h]
% \vspace{-0.5cm}
    \centering
    \includegraphics[width=0.3\textwidth]{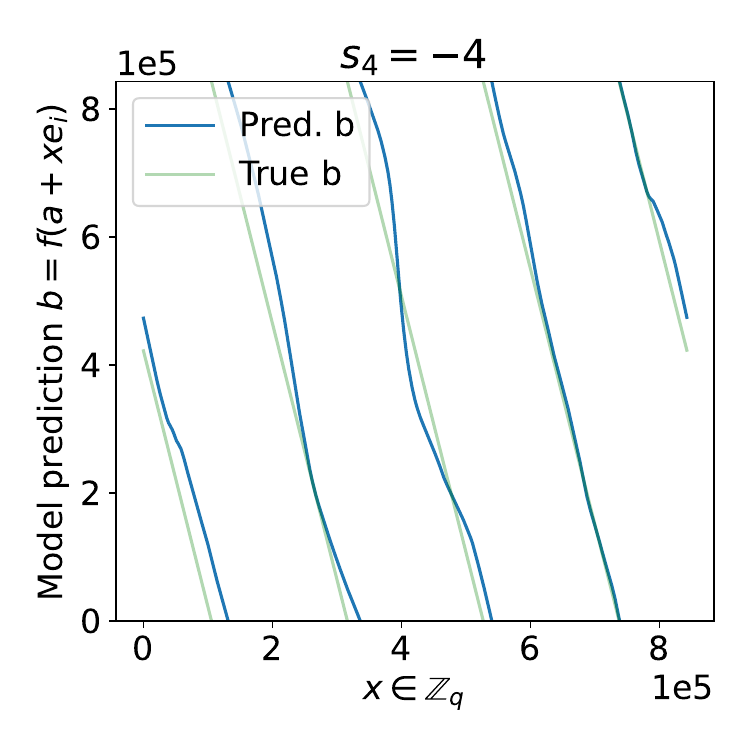}
    \caption{{\bf Slope distinguisher for recovering general secrets.} This distinguisher computes $b=f({\bf a} + x {\bf e_i})$ for varying $x \in [0,q]$ and recovers secret bit values from the slope of this line. This plot is for $s_4 = -4$. The blue line ``pred b'' plots model outputs $b=f({\bf a} + x {\bf e_i})$ for $x \in [0,q]$ for some fixed in-distribution $a$. The green line ``true b'' shows $f_{\text{true}}({\bf a} + x {\bf e_i}) = {\bf a}\cdot {\bf s}+xs_i$. Model $f$ is trained on BKZ-reduced LWE data with Gaussian secrets, $n=256, \log q =20$. }
    \label{fig:model_output}
    \vspace{-0.2cm}
\end{figure}

\vspace{0.1cm}
\noindent {\bf {\em Recovering general secrets.}} ~\cite{salsa} recovers binary secrets by observing whether the output of the model $f$ changes when input values are modified at a particular index $i$. Since secret bits are binary, this signal is sufficient. Recovering general secrets (like binomial), however, requires also finding the value of the active secret bits. To do this, one might consider modifying input elements using the vector $\delta \bf{e_i}$, where $\delta$ is a small value and $\bf{e_i}$ is the standard basis vector which is $1$ at the $i$-th component and $0$ everywhere else. When $f$ encounters this input, one would expect it to output $b\approx \delta s_i \mod q$, revealing this secret bit value. 

However, we observe experimentally that since $\delta \bf{e_i}$ falls outside of $f$'s training distribution, $f$ does not produce helpful predictions on this input. Thus, we propose an alternative secret recovery method that embodies this concept, while remaining within the data distribution. We call it a ``slope distinguisher.'' This distinguisher calculates the slopes, or approximations of the derivatives, of model outputs using the formula $\frac{\partial f}{\partial x_i}(a) \approx \frac{f(a+\delta e_i) - f(a)}{\delta} =: \hat{s}_i$ for some $a$ drawn from the data distribution. To account for the noisy predictions, we compute many samples of $\hat{s}_i$ and take the most frequently appearing value, rounded to the nearest integer. With this, the ML method can recover general secrets. This new distinguisher is  illustrated in \cref{fig:model_output}.

\vspace{0.1cm}
\noindent {\bf {\em Data preprocessing and model training.}} We follow the interleaved reduction strategy of~\cite{stevens2024fresca} and use both \texttt{flatter} and \texttt{BKZ2.0} for data preprocessing. We preprocess $\bf A$ matrices until reduction factor $\rho = \frac{\sigma(\mathbf{RA})}{\sigma(\mathbf A)}$ flatlines, where $\sigma$ denotes the mean of the standard deviations of the rows of $\mathbf{RA}$ and $\mathbf{A}$. Table~\ref{tab:ml_cc_preproc} gives $\rho$ for each setting. For most settings, LWE matrices have $m=0.875$n, but for larger $n$ with smaller $q$, we find using $m>n$ enables better reduction. For the ($k=2$, $\log_2q=12$) \kyb{} setting, we use $m=712$, and for the ($\log_2q = 26, 29$) HE settings, we use $m=1624$. All others use $m=0.875n$. We set $\omega=10$ for the HE benchmark datasets and $\omega = 4$ for the \kyb{} datasets. This is because the $\binomsec$ error with $\eta=2$ is smaller than $\normsec$ with $\sigma=3$, and so a smaller $\omega$ can be tolerated. Again following~\cite{stevens2024fresca}, we create datasets of $2$ million LWE examples and use an encoder-only transformer with an angular embedding, which represents integers mod $q$ as points on the unit circle. We do not pre-train our transformers, but train each one fresh on a dataset with unique secret $\mathbf{s}$ on one GPU. 

\begin{table}[h]
\resizebox{0.49\textwidth}{!}{
\begin{tabular}{cccc@{\hskip 0.4in}ccc}
\toprule
%\multirow{2}{*}{\textbf{\begin{tabular}[c]{@{}c@{}}Setting\\ ($k, \log_2 q$)\end{tabular}}}
{\bf Setting} & \multicolumn{3}{c}{\kyb{} ($n=256$)} & \multicolumn{3}{c}{HE ($n=1024$)} \\ \cmidrule{2-7}
($k, \log_2 q$) & ($2, 12$) & ($2, 28$) & ($3, 35$) & ($1, 26$) & ($1, 29$) & ($1, 50$) \\ \midrule
$m$ & $712$ & $448$ & $672$ & 1624 & 1624  & 896 \\
\textbf{$\rho$} & 0.88  & 0.67 & 0.69 & 0.86 & 0.84 & 0.70 \\
\# cruel bits & 388 & 228 & 381 & 750 & 715 & 495 \\ 
reduction time (hrs) & 27.7 & 10.5 & 23.3 & 21.5 & 31.6 & 23.8 \\
\# samples/matrix. & 621 & 664 & 1084 & 1725 & 1717 & 1558 \\

\bottomrule
\end{tabular}
}
\caption{{\bf Data preprocessing for ML and CC attacks}. $\rho$ measures the overall standard deviation reduction of $\bf RA$, relative to $\bf A$. \# cruel bits = number of unreduced bits in CC attack. Reduction time = hours needed to reduce a matrix to the given $\rho$ and \# cruel bits. \# samples/matrix = the average number of reduced LWE samples extracted per matrix when reducing the embedded matrix of shape $(m+n)\times (m+n)$. This number is less than $m+n$ because we discard rows of $R$ which are all $0$.}
\label{tab:ml_cc_preproc}

\end{table}

%\vspace{-0.3cm}
\subsection{Cool \& Cruel Attack}
\vspace{-0.2cm}

Next, we implement the ``cool and cruel'' attack of~\cite{nolte2024cool}. The first part of this attack is the data preprocessing step of the ML attack described above: starting with $4n$ LWE samples, run lattice reduction on LWE matrices subsampled from these to produce a large dataset of reduced LWE samples (see prior section for details). After data preprocessing, the cruel and cool bits are identified by inspecting the standard deviations of the columns of $\mathbf{RA}$. Columns with standard deviation $\sigma$ greater than $\frac{q}{2\sqrt{12}}$, assuming the original $\mathbf{A} \sim \mathbb{U}(0,q)$, are ``cruel'' and the rest are ``cool.'' Cool bits can be ignored during the first part of secret recovery. Their norm is so small that their contribution to $\mathbf{Rb}$ is minimal. If the cruel bits are correctly guessed (via brute force), the residuals $r = \mathbf{Rb} - \mathbf{RA}\cdot s_{cruel}$ have a distribution closer to normal than uniform random, which can be statistically detected. After cruel bits are recovered, cool bits are recovered greedily. 

\para{Implementation details.} We use the RLWE ``cliff-shifting'' trick for secret recovery in the HE parameter regime, and adapt the MLWE ``split-cliff-shifting'' described in~\ref{subsec:ml} for Kyber. Additionally, the attack must be adapted to recover ternary, binomial and Gaussian secrets, since the original paper only considers binary secrets. Brute force recovery of cruel bits is unaffected by a change in secret distribution (although the number of guesses increases exponentially), but we find experimentally that the cool bit recovery of Algorithm 1 of~\cite{nolte2024cool} fails on wider secret distributions.

\vspace{0.1cm}
\noindent {\bf {\em Linear regression method.}} We develop a linear regression-based method to recover cool bits in wider secret distributions. The underlying rationale is that once cruel bits are recovered, the remaining elements of $a$ are sufficiently small to prevent most of the residual dot products from wrapping around $q$. Hence, linear regression could be used. This method works as follows. Consider $\mathbf A' = \mathbf{RA} = \begin{pmatrix} \mathbf A_u' & \mathbf A_r' \end{pmatrix}$ where $\mathbf A_u'$ are the un-reduced entries of $\mathbf{A}'$ and $\mathbf A_r'$ are the reduced entries. Then $\mathbf Rb \underset{\mathbf Re}{\approx} \mathbf{A}'s \mod q = \begin{pmatrix}\mathbf A_u' & \mathbf A_r' \end{pmatrix} \cdot \begin{pmatrix} s_u & s_r\end{pmatrix}^{\top} = \mathbf A_u's_u + \mathbf A_r's_r$. If $s_u$ is known, e.g. through brute force, then the linear regression applied to the pair $(X,y) = (\mathbf A_r', \mathbf Rb - \mathbf A_u' s_u \mod q)$  (where$\mod q$ centers entries to $(-\frac{q}{2}, \frac{q}{2})$) yields a least-squares estimator for $s_r$: $\hat{s}_r = (\mathbf A^{'\top}_r \mathbf A_r')^{-1}\mathbf A_r^{'\top}(\mathbf Rb - \mathbf A_u' s_u \mod q)$. Using this approach, we recover ternary, binomial, and Gaussian secrets.

\begin{table*}[t]
\centering

\begin{tabular}{clccc@{\hskip 0.4in}ccc} \toprule
 & & \multicolumn{3}{l}{\kyb{} Benchmark Setting} & \multicolumn{3}{c}{RLWE Benchmark Setting} \\ \midrule
Parameters & $(n, k)$ & (256, 2) & (256, 2) & (256, 3) & (1024, 1) & (1024, 1) &  (1024, 1) \\
 & $\log_2 q$ & 12 & 28 & 35 & 26 & 29  & 50 \\
 & $\tau$ & 50 & 50 & 50 & 50 & 50 & 50 \\
 & $\zeta$ & 500 & 325 & 540 & 920 & 828  & 650 \\ \midrule
 Error bound & $B/q$ & 0.11 & 0.04 & 0.02 & 0.02 & 0.04 & 0.02 \\ \midrule
\multirow{4}{*}{\begin{tabular}[c]{@{}c@{}}MiTM table size \\ for varying $h'$\end{tabular}} & $h'=4$ & 10 MB & 10 MB & 10 MB & 30 MB & 30 MB &  30 MB \\
  & $h'=6$ & 2.0 GB & 0.5 GB & 2.5 GB& 12.2 GB & 8.9 GB & 7.5 GB \\
 & $h'=8$ & 244 GB & 43 GB& 331 GB & 2.8 TB & 1.8 TB & 1.2 TB\\
 & $h'=10$ & $28$ TB & $3.3$ TB & $42$ TB & $600$ TB & $354$ TB & $173$ TB \\ \bottomrule
\end{tabular}
\caption{{\bf Experimental parameters and estimated memory requirements for our implementation of the MiTM attack on Decision-LWE.} $\tau$ = \# of short vectors used for the guessing step, $\zeta$ = the guessing dimension. $h'$ = \# of nonzero secret bits in the $\zeta$ guessing region. }%Error bound $B/q$ is computed using the definition of~\cite{Cheon_hybrid_dual}: $B = (2 + \frac{1}{\sqrt{2\pi}}) \cdot \alpha q \sqrt{\frac{m}{m+n}} \cdot \frac{||\mathbf{y_1}||}{c}$. We use average norm of short vectors for a setting as $||\mathbf{y_1}||$.}
\label{tab:both_mitm_mem}
\vspace{-0.3cm}
\end{table*}

\vspace{0.1cm}
\noindent {\bf {\em Data preprocessing.}} We follow the same strategies and parameters as described in \S\ref{subsec:ml}. Table~\ref{tab:ml_cc_preproc} gives the number of cruel bits per setting after reduction. Brute force recovery runs on GPUs and can be parallelized by dividing up the set of Hamming weights to be guessed. We use $5K$ reduced LWE samples for the brute force portion of the attack, and $100K$ samples for the Linear Regression recovery. For parity with other attacks, we use one GPU per experiment.

\vspace{-0.2cm}
\subsection{Dual Hybrid MiTM}
\label{subsec:mitm}

Finally, we consider the dual hyrid MiTM, which attacks decision-LWE, not search-LWE. Although this attack does not actually recover secrets, it is relevant because of its focus on low $h$ secrets.  We base our implementation of dual hybrid MiTM on~\cite{Cheon_hybrid_dual} since they provided code.

The attack works as follows. Given LWE samples ($\mathbf{A} \in \mathbb{Z}^{m \times n}_q, \mathbf{b} \in \mathbb{Z}^m_q$), choose a guessing dimension $\zeta$, and split $A$ along this. This creates $\bf A = A_1 || A_2$ with $\mathbf{A_1} \in \mathbb{Z}_q^{m \times (n-\zeta)}$ and $\mathbf{A_2} \in \mathbb{Z}_q^{m \times \zeta}$  and implicitly divides the secret into $\bf s = s_1 || s_2$, corresponding to $\bf A_1, A_2$, and $\bf b$ into $\bf b = A_1 \cdot s_1 + A_2 \cdot s_2 + e$. Then, create a scaled normal dual lattice $\Lambda_{q,c}(\mathbf{A_1}) = \{(\mathbf{v}_1, \mathbf{v}_2 \in \mathbb{Z}^m \times (\frac{1}{c}\mathbb{Z}^n) : \mathbf{v}_1^t \mathbf{A_1} \equiv_q c \cdot \mathbf{v}_2\}$\cite{bai_galbraith}, and run lattice reduction on $\Lambda_{q,c} (\mathbf{A_1})$ to find short vectors $(\mathbf{y}_1, \mathbf{y}_2) \in \Lambda_{q,c}(\mathbf{A_1})$. These short vectors can then be applied to $\bf b$ to minimize the contribution of $\mathbf{s}_1$: 
\begin{align*}
    \langle \mathbf{y}_1, \mathbf{b} \rangle &= \langle \mathbf{y}_1, \mathbf{A_1}\mathbf{s}_1 \rangle + \langle \mathbf{y}_1, \mathbf{A_2} \mathbf{s}_2 \rangle + \langle \mathbf{y}_1, {\bf e}\rangle \\ 
    & \equiv_q \mathbf{y}^t_1 \mathbf{A_2}\mathbf{s}_2 + c \cdot \mathbf{y}^t_2\mathbf{s}_1 + \mathbf{y}^t_1\mathbf{e}
\end{align*}

\noindent If $(\mathbf{y}_1, \mathbf{y}_2)$ are sufficiently short, then one can simply consider $c \cdot \mathbf{y}^t_2\mathbf{s}_1 + \mathbf{y}^t_1\mathbf{e}$ as a new error term $\mathbf{e}'$. This creates a new LWE sample ($\mathbf{A', b'}) \in \mathbb{Z}_q^{m \times \zeta}, \mathbb{Z}_q^m$, where $\mathbf{A'} = \mathbf{y}^t_1\mathbf{A_2}$ and $\mathbf{b'} = \mathbf{y}^t_1\mathbf{A_2}\mathbf{s}_2 + \mathbf{e'}$. This step must be repeated $\tau$ times to generate a sufficient number of reduced LWE samples to construct the MiTM table and guess $\mathbf{s}_2$. 

The MiTM approach guesses the components of $\mathbf{s}_2$ using a lookup table. It first constructs a table $\mathcal{T}$ holding all possible secret candidates with $h_1 \le h = |\mathbf{s}|$ (e.g. $h$ is the number of nonzero coordinates of $\mathbf{s}$). $\mathcal{T}$ is indexed by a locality-sensitive hash function that operates on a secret candidate $\mathbf{s}^*$ as follows. Compute $\mathbf{b}^* = \mathbf{A}'\mathbf{s}^* \in \mathbb{Z}_q^{\tau}$ and create zero string $I = {0}^{\tau}$. For each element of $\mathbf{b}^*$, let $I_i = 1$ if $\mathbf{b}^*_i < q /2$, else $0$, then set $\mathcal{T}[I] = \mathbf{b}^*$.  Using $\mathcal{T}$, the goal is to guess a secret $\mathbf{s}^{\dagger}$ such that $r^{\dagger} = \mathbf{b}' - \mathbf{A}'\mathbf{s}^{\dagger} = \mathbf{A}' \mathbf{s}^*$ is in $\mathcal{T}$. If this collision occurs, $\mathbf{s}_2' = \mathbf{s}^{\dagger} || \mathbf{s}^*$ is a possible secret candidate and can be quickly checked for correctness.  %

An important parameter of MiTM is the error bound $B$.
During MiTM, $B$ calibrates the sensitivity of the locality-sensitive hash. If an element of $r^{\dagger}$ falls in the range $[0, B)$, $(q-B, q]$, or ($q/2 - B, q/2 + B)$, the error introduced in creating the new LWE sample could have ``flipped'' this element around the modulus. Thus, one must recursively flip each hash index associated with a boundary element, to ensure the true secret candidate is not missed. Search time increases exponentially in $\tau$, the length of the hash string, and $B$: $\mathcal{O}(2^{4\tau B/q})$.

\begin{table*}[]
%\parbox{.45\linewidth}{
%\resizebox{1.0\textwidth}{!}{
\centering 
\begin{tabular}{cl@{\hskip 0.3in}ccc@{\hskip 0.3in}ccc}
\toprule  

\multirow{3}{*}{Attack} & \multirow{3}{*}{Results} & \multicolumn{3}{c}{Kyber MLWE Setting ($n$, $k$, $\log_2 q$)} & \multicolumn{3}{c}{HE LWE Setting ($n$, $\log_2 q$)} \\ 
\cmidrule{3-8}
%\begin{tabular}[c]{@{}c@{}}LWE Setting\\ ($n$, $\log_2 q$)\end{tabular} &  & (512, 12) & (512, 28) & (768, 35) \\ \hline
 &  & (256, 2, 12) & (256, 2, 28) & (256, 3, 35) & (1024, 26) & (1024, 29)  & (1024, 50)\\
 &  & binomial & binomial  & binomial & ternary & ternary & ternary \\
 \midrule
\multirow{2}{*}{uSVP} & Best $h$ & -  & - &  -&-  &-  &-  \\
 & Recover hrs (1 CPU)   & $>1100$ & $>1100 $ & $>1300 $  &  $>1300 $ & $>1300 $  & $>1300$ \\ \midrule
% & Recovery time & - & - & - & - & - & - & - & -\\
% & Success rate  & - & - & - & - & - & - & - & -\\
% & Memory (MB) & $\approx 50$ & $\approx50$ & $\approx50$ & $\approx50$  & $\approx50$ & $\approx50$ & $\approx50$ & $\approx50$\\ \midrule
 
\multirow{4}{*}{ML} & Best $h$ & 9 & 18 & 16 & 10 & 10 & 17  \\
 & Preproc. hrs $\cdot$ CPUs & $28$  $\cdot$ $3216$ & $11$ $\cdot$ $3010$ & $33$ $\cdot$ $1843$   & $32$ $\cdot$ $1160$ &$31.6$ $\cdot$ $1164$   & $23.8$ $\cdot$ $1284$\\ 
 & Recover hrs $\cdot$ GPUs & $8$ $\cdot$ $256$ & $16$ $\cdot$ $256$ & $6$ $\cdot$ $256$  & $14 \cdot 1024$ & $17.8 \cdot 1024$  & $5.3 \cdot 1024$ \\
 & Total hrs & $36$ & $27$ & $39$ & $46$ & $49.4$ &  $29.1$ \\ \midrule
 %& Success rate & $1/10$ & $1/10$ & $1/10$ & $1/10$ & $1/10$ &  &  & $1/10$ \\
 %& Memory &  &  &  \\ \midrule
 
\multirow{4}{*}{CC} & Best $h$ & {\bf 11} & {\bf 25} & {\bf 19} & \bf{11} & \textbf{12} & {\bf 20} \\
 & Preproc. hrs $\cdot$ CPUs & { 28 $\cdot$ 161} & { 11 $\cdot$ 151} & { 23 $\cdot$ 92} & $32$ $\cdot$ $58$ & { 31.6 $\cdot$ 58} & { 23.8 $\cdot$ 64}
 \\
 & Recover hrs $\cdot$ GPUs & { 0.1 $\cdot$ 256} & { 42 $\cdot$ 256} & { 0.9 $\cdot$ 256}  & $0.1 \cdot 1024$ & { 0.1 $\cdot$ 1024} & { 4.2 $\cdot$ 1024} \\ 
 & Total hrs & { 28.1} & { 53} & { 34} & 32.1 & { 31.7}  & { 28} \\ \midrule \midrule
% & Success rate & $2/10$ & $1/10$ & $2/10$ & $2/10$ & $1/10$ &  &  & $2/10$ \\
% & Memory &  &  & &  &  &  &  &  \\ \midrule
 % \begin{tabular}[c]{@{}c@{}}MiTM\\(Decision \\LWE)\\\end{tabular}
\multirow{4}{*}{\begin{tabular}[c]{@{}c@{}}MiTM\\(Decision\\LWE)\end{tabular}} & Best $h$ & 4 & 12 & 14 & 9 & {9} & 16 \\
 & Preproc. hrs $\cdot$ CPUs  & $0.5$ $\cdot$ $50$ & $1.6$ $\cdot$ $50$ & $4.4$ $\cdot$ $50$ & { 8 $\cdot$ 50} & { 11.4 $\cdot$ 50} & $14.4$ $\cdot$ $50$  \\
% & Multi-CPU hrs $\cdot$ CPUs &  $\cdot$&  & &  &  &  &  &  \\
 & Decide hrs (1 CPU) & 0.2 & 0.01 & 25 & { 57} & { 2} & 1.1 \\  
 & Total hrs & 0.7 & 1.61 & 29.4 & { 65} & { 13} & 15.5 \\  \bottomrule 
% & Success rate  &  &  &  &  &  &  &  & \\
% & Memory  &  &  &  &  &  &  &  & \\ \bottomrule
\end{tabular}
%}
\caption{{\bf  Performance of all attacks on benchmark settings.} Best Hamming weight $h$ for secret recovered per setting/attack, time in hours needed to recover this secret, and machines used. Highest $h$ per setting is {\bf bold}. All Kyber secrets are binomial, and HE secrets are ternary. First three attacks (uSVP, ML, CC) are Search-LWE; MITM* is Decision LWE. The ML, CC, and MiTM attacks have two phases: Preprocessing (Prepoc. in table), when LWE data is reduced and/or short vectors are obtained; Recovery (Recover in table) for ML/CC, when reduced vectors are used recover secrets; and Decide for MiTM, when Decision LWE is solved using short vectors. We report time separately for each step. When steps can be parallelized, we report hours/machine and number of machines. The uSVP attack has only the ``recover'' phase, which cannot be parallelized. ``Total hrs'' is total attack time assuming full parallelization.}% For attacks where parallelization is possible (all but uSVP), we report multi- and single-core times, and number of machines used. 
%\caption{{\bf All benchmark results.} All times are in hours. Categories presented per attack: best hamming weight, preprocessing time, recovery time, probability of success, memory used (CC/AI = dataset memory + 32GB GPU). '-' indicates that an attack did not succeed. \todo{decide what to do with ternary vs gaussian secrets.} \todo{add summed time} \todo{bold best attack at each setting}}
\label{tab:all_results}
\vspace{-0.2cm}
\end{table*}

\para{Implementation details.} Our implementation builds on that of~\cite{Cheon_hybrid_dual} but makes several improvements to enable the first known evaluation of dual hybrid MiTM on LWE problems with $n>100$.  We check $\mathbf{s}^{\dagger}$ candidates {\em as $\mathcal{T}$ is created}\textemdash every time we insert a new $\mathbf{s}^{\dagger}$ candidate into $\mathcal{T}$, we also check if $r^{\dagger} = \mathbf{b}' - \mathbf{A}' \mathbf{s}^{\dagger}$ is in $\mathcal{T}$. We combine BKZ2.0 with $\beta=30$ and \texttt{flatter} for the scaled dual attack to improve efficiency. We expand the attack to include RLWE and MLWE settings, as well as ternary, binomial, and Gaussian secrets. Since $|\mathcal{T}|$ grows exponentially with possible secret bit values, we trade off memory and time by storing indices of possible nonzero secret bits in $\mathcal{T}$ (e.g. $[0, 46, 127]$), and exhausting over secret bit values for each guess (e.g. $[1,1,1], [1,1,-1], \ldots [-1,-1,-1]$ for $\ternsec$). Finally, we assume the attacker has $\tau n$ initial samples.

\vspace{0.1cm}
\noindent {\bf {\em Parameters.}} The definitions of $\tau$, $B$ and $c$ given on~\cite[p.21]{Cheon_hybrid_dual} depend on the root Hermite factor $\delta_0$ of the short dual vectors, and a target value for $\delta_0$ is not provided. Hence, we make the following engineering choice, based on experiments. We observe that only short vectors with $B < Q/8$ result in MiTM searches that run in reasonable time, so we fix $B < q/8$ and compute it using the method of the~\cite{Cheon_hybrid_dual} implementation: $B = (2 + \frac{1}{\sqrt{2\pi}}) \cdot \alpha q \sqrt{\frac{m}{m+n}} \cdot \frac{||\mathbf{y_1}||}{c}$. $\mathbf{y_1}$ is a short vector obtained from the scaled dual attack, and we use the average norm of all short vectors obtained from the scaled dual reduction to compute $B$. The definition of $B$ in the paper relies on $\delta_0$, but formulae for estimating $\delta_0$ are inaccurate for $\beta < 50$~\cite{chen2013reduction, CCLS}. We fix $c=10$, mirroring~\cite{Cheon_hybrid_dual}, and use $m=n$, following~\cite{Cheon_hybrid_dual, albrecht2017lattice}.

For $\tau$ and $\zeta$, we initially use values provided by the Lattice Estimator but find experimentally that these are inaccurate. For example, $\tau=50$ short vectors are sufficient to recover secrets, compared to the hundreds estimated by Estimator. We also find that $\zeta$ values given by the estimator make the reduction of these dual lattices unreasonably slow for the small block sizes we can tractably run. We run ablation experiments across various $\zeta$ and $\beta$ values for the other two settings, and use $\zeta$ values providing the best trade off in reduction time and secret recovery. Table~\ref{tab:both_mitm_mem} lists our chosen $\zeta$ and $\tau$ and experimentally chosen error bound $B/q$. 
\vspace{0.1cm}
\noindent {\bf {\em Search criterion.}} In the~\cite{Cheon_hybrid_dual} implementation, the correctness of a table element is assessed by computing the $L_{\inf}$ norm (e.g. largest element) of the putative short vector $\mathbf{r}' = \mathbf{b}' - \mathbf{A}'s^* - \mathbf{A}'s^{\dagger}$, where $\mathbf{s}^*$ is an element stored in the table and $\mathbf{s}^{\dagger}$ is a guess. However, we observe that $\mathbf{r}'$ often contains outliers, so using the $L_{\inf}$ norm may yield false negatives. We instead check against the {\em median} value of $\mathbf{r}'$, which reduces the effect of large outliers. % and correctly recovers true secret guesses.

\vspace{0.1cm}
\noindent {\bf {\em Memory Constraints.}}  MiTM memory requirements scale exponentially with secret Hamming weight.
In Table~\ref{tab:both_mitm_mem} we provide the memory requirements for implementing the table look-up for 
Hamming weight $h'$ with a guessing region of length $\zeta$.
This shows what secrets are recoverable on our hardware, with $750$ GB RAM.  So we can recover secrets with $h' \le 8$ for all Kyber settings, and $h' \le 6$  for $\log_2q \le 34$ and $h' \le 8$ for $\log_2 q = 45, 50$ (if search time is $< 72$ hours, our computer cluster limit). One can then compute the probability that a secret with Hamming weight $h$ has this $h'$ value, and use this to estimate which $h$ secrets are recoverable\textemdash see Appendix \S\ref{appx:mitm_probs} for more details.

\vspace{0.1cm}
\noindent {\bf {\em Dual Hybrid MiTM solves Decision-LWE, not Search-LWE.}} Solving search-LWE would require some additional solution and implementation and add to the total cost. All other benchmarked attacks solve Search-LWE.

\section{Measuring Attack Performance}
\label{sec:benchmark}
\vspace{-0.2cm}

Having implemented and improved these four attacks, we now evaluate them on our proposed settings. Table~\ref{tab:all_results} records best results for all attacks across all settings, using the evaluation metrics of \S\ref{subsec:metrics}. All attacks are run on the same randomly generated secrets. For each setting, we {\bf bold} the highest $h$ recovery. Tables~\ref{tab:kyber_detailed_all} and~\ref{tab:ml_he_practical} give detailed results for experiments on the \kyb{} and HE benchmark settings, reporting the success rate of attacks for a range of Hamming weight secrets per setting and the best time (in hours) for recovering a secret at each $h$. 

\begin{table*}[t]
\centering
\resizebox{1.0\textwidth}{!}{
\begin{tabular}{llllllllcllll}
\toprule
\multirow{2}{*}{($n$, $k$)} & \multirow{2}{*}{$\log_2 q$} & \multirow{2}{*}{$h$} & \multicolumn{3}{c}{USVP (Search-LWE)} \hspace{0.4cm} & \multicolumn{7}{c}{Dual Hybrid MITM (Decision-LWE)} \\ \cmidrule{4-13}
 &  &  \hspace{0.3cm} & ROP & time (yrs) & BKZ $\beta$ \hspace{0.4cm} & ROP & repeats & single time / time w. repeats  & memory & $\tau$ & $\zeta$ & $h'$ \\ \midrule

 % 512 12
($256$, $2$) & 12 & 4 & $2^{260.0}$ & $2.8e61$ & 382 & $2^{33.7}$ & $2^{7.2}$  & $6$ secs / 16 mins & $2^{22.3}$ (0.02 GB)  & 92& 476 & 3 \\ % 16 mins w repeats

 % 512 28
($256$, $2$) & 28 & 12 & $2^{62.8}$ & $120$ & 109 & $2^{40.6}$ & $2^{8.3}$ & $13$ mins / 68.6 hrs & $2^{30.6}$ (6.7 GB) & 310 & 308 & 6 \\ % 68.6 hrs w/ repeats

 % 768 35
($256$, $3$) & 35 & 14 & $2^{81.7}$ &$5.9e7$ & 142 & $2^{43.2}$ & $2^{9.0}$ & $1.4$ hrs / 29.4 days & $2^{32.6}$ (27 GB) & 441 & 444 & 6 \\ \midrule % 29.4 days w/ repeats

% logq 26 with SPARSE TERNARY
($1024$, $1$) & 26 & 9 & $2^{203.5}$ & 2.55e44 & 313 & $2^{42.6}$ & $2^{11.4}$ &  $0.9$ hrs / $106$ days & $2^{29.7}$ (3.5 GB) & 251 & 887 & 5 \\ % $106$ days with repeats

 % logq 29
($1024$, $1$)  & 29 & 9 & $2^{244.9}$ & 8.1e56 & 363 & $2^{42.2}$ & $2^{9.6}$ &  $0.7$ hrs / $21.6$ days&  $2^{31.7}$ (14 GB) & 297 & 823 & 5 \\ % $21.6$ days with repeats

 % logq 34
%($1024$, $1$)  & 34 & 9 & $2^{170.7}$ & 3.8e34 & 271 & $2^{41.7}$ & $2^{8.4}$ & $7$ days & $2^{32.6}$ (110 GB) & 372 & 748 & 5 \\

 % logq 45
%($1024$, $1$)  & 45& 11 & $2^{100.6}$ & 3.0e13 & 172 & $2^{42.6}$ & $2^{8.2}$ & $11.5$ days & $2^{33.7}$ (264 GB) & 480 & 667 & 6 \\

 % logq 50
($1024$, $1$)  & 50 & 12 & $2^{83.4}$ & 1.9e8 & 144 & $2^{44.8}$ & $2^{9.0}$ & $4$ hrs / $85$ days &$2^{33.6}$ (103 GB) & 620 & 523 & 6 \\ \bottomrule %$7$ days  with repeats
 
\end{tabular}
}
\caption{{\bf Estimated performance of uSVP and Dual Hybrid MiTM attacks on Kyber and HE benchmark settings using Chen Nguyen cost model~\cite{CN11_BKZ}} We modify the Estimator to consider blocksize $\beta \ge 20$, instead of default $\ge 40$, to better estimate performance. According to Estimator documentation, ``ROP'' approximates the number of CPU cycles needed to run the attack, so we convert ROP to time by dividing this by 2.1GHz ($2.1e9$ cycles/sec), the clock speed of our CPUs. For Dual Hybrid, we present time and time multiplied by estimated repeats (number of times attack should run to succeed with probability 0.99). We convert predicted memory to bytes by multiplying estimator output (number of integers to be stored) by number of bits needed to store integer (based on $\log_2 q$).} % For Dual Hybrid, we multiply by estimated repeats, the number of times the attack must run to succeed with probability $0.99$.}
%\vspace{-0.3cm}
\label{tab:kyber_he_estimates}
\end{table*}

\begin{table*}[t]
\begin{minipage}{0.49\textwidth}
\centering
\resizebox{0.99\textwidth}{!}{
\begin{tabular}{cccc@{\hskip 0.3in}ccc@{\hskip 0.3in}ccc}
\toprule
\multirow{2}{*}{Attack} & 
\multicolumn{3}{c}{\hspace{-0.2in} $k=2,\;\log q=12$} & \multicolumn{3}{c}{\hspace{-0.2in} $k=2,\;\log q=28$} & \multicolumn{3}{c}{ \hspace{-0.1in} $k=3,\;\log q=35$} \\[0.1cm]
& $h$ & Rate & Time & $h$ & Rate & Time & $h$ & Rate  &  Time \\
\midrule
\multirow{4}{*}{ML} & 
6 & 6 / 10 & $1.2$ & 15 & 5 / 10& $1.2$ & 14 & 2 / 10& $1.2$ \\
& 7 & 1 / 10 & $1.6$ & 16 & 2 / 10& $1.7$ & 15 & 3 / 10& $15.8$ \\
& 8 & 0 / 10 & - & 17 & 1 / 10& $25.7$ & 16 & 1 / 10& $3.7$ \\
& 9 & 1 / 10 & $7.4$ & 18 & 1 / 10& $17$ & 17 & 0 / 10 & - \\
\midrule
\multirow{5}{*}{CC} & 
7 & 10 / 10 & $0.03$ & 19 & 4 / 10 & $0.1$ & 16 & 4 / 10 & $0.03$ \\
& 8 & 7 / 10 & $0.06$ & 20 & 3 / 10 & $0.2$ & 17 & 6 / 10 & $0.1$ \\
& 9 & 6 / 10 & $0.04$ & 21 & 3 / 10 & $0.6$ & 18 & 4 / 10 & $0.03$ \\
& 10 & 1 / 10 & $1.4$ & 24 & 2 / 10 & $1.0$ & 19 & 2 / 10 & $0.9$ \\
& 11 & 2 / 10 & $0.1$ & 25 & 1 / 10 & $41.8$ & 20 & 0 / 10 & - \\
\midrule
\midrule
\multirow{3}{*}{MiTM} & 
4 & 2 / 10 & 0.2 & 11 & 1 / 10& 0.14 & 12 & 1 / 10& 19 \\
& 5 & 0 / 10 & - & 12 & 1 / 10&  0.02 & 13 & 2 / 10&   15  \\
& 6 & 0 / 10 & - & 13 & 0 / 10& - & 14 & 1 / 10& 25 \\
%9 & 1 / 10 & $2^{22.7}$ & 18 & 1 / 10& $2^{23.9}$ & 17 & 0 / 10 & - \\
\bottomrule
\end{tabular}
}
\caption{{\bf Detailed attack results on Kyber benchmark settings for varying $h$, $n=256$ for all, binomial secrets.} Rate = secrets recovered / attempted. Time = in hours, best single CPU/GPU time for recovering secrets via model training/brute force/MiTM table queries. Required compute resources are given in Table~\ref{tab:all_results}. Preprocessing/short vector generation time is the same for all $h$ at a given setting and is listed as Preproc. hrs in Table~\ref{tab:all_results}. }
\label{tab:kyber_detailed_all}
\end{minipage}\hfill
\begin{minipage}{0.49\textwidth}
\resizebox{0.99\textwidth}{!}{
\begin{tabular}{cccccccccc}
\toprule
\multirow{2}{*}{Attack} &  \multicolumn{3}{c}{$\log q=26$} & \multicolumn{3}{c}{$\log q=29$} & \multicolumn{3}{c}{$\log q=50$} \\[0.1cm]
%& \multicolumn{3}{c}{ternary} & \multicolumn{3}{c}{gaussian} & \multicolumn{3}{c}{gaussian} & \multicolumn{3}{c}{gaussian} & \multicolumn{3}{c}{gaussian} \\ \midrule
& $h$ & Rate & Time  & $h$ & Rate & Time & $h$ & Rate & Time \\ \midrule
\multirow{4}{*}{ML} & 7 & 3 / 10 & $3.5$ & 8 & 1 / 10 & $15.9$ & 14 & 1 / 10 & $3.9$ \\
 & 8 & 0 / 10 & - & 9 & 2 / 10 & $3.1$ & 15 & 0 / 10 & - \\
 & 9 & 0 / 10 & - & 10 & 1 / 10 & $17.8$ & 16 & 2 / 10 & $20.4$ \\
 & 10 & 1 / 10 & $14$ & 11 & 0 / 10 & - & 17 & 1 / 10 & $5.3$ \\ \midrule
\multirow{4}{*}{CC}  & 9 & 8 / 10 & 0.07 & 9 & 10 / 10 & $0.03$ & 17 & 6 / 10 & $0.05$ \\
 & 10 & 3 / 10 & 0.07 & 10 & 0 / 10 & - & 18 & 4 / 10 & $1.9$ \\
 & 11 & 1 / 10 & 0.1 & 11 & 0 / 10 & - & 19 & 6 / 10 & $0.07$ \\
 & 12 & 0 / 10 & - & 12 & 1 / 10 & $0.13$ & 20 & 2 / 10 & $4.2$ \\ \midrule \midrule
\multirow{3}{*}{MiTM} & 7 & 4 / 10 & 0.2 & 7 & 1 / 10 & 1 & 14 & 1 / 10 & 0.4 \\
 & 8 & 2 / 10 & 38 & 8 & 0 / 10 & - & 15 & 0 / 10 & - \\
 & 9 & 1 / 10 & 57 & 9 & 2 / 10 & 2 & 16 & 1 / 10 & 1.1 \\ \bottomrule
\end{tabular}
}
\caption{\textbf{Detailed attack results on HE benchmark settings for varying $h$, $n=1024$ for all settings, ternary secrets.} Rate = secrets recovered / attempted. Time = in hours, best single CPU/GPU time for recovering secrets via model training/brute force/MiTM table queries. Required compute resources are given in Table~\ref{tab:all_results}. Preprocessing/short vector generation time is the same for all $h$ at a given setting and is listed as Preproc. hrs in Table~\ref{tab:all_results}.}
\label{tab:ml_he_practical}
\end{minipage}
\end{table*}

\para{Preprocessing/Recover/Decide Time.} Table~\ref{tab:all_results} lists ``preproc'', ``recover'', and ``decide'' times, along with compute requirements. These refer to the distinct steps in the ML, CC, and MiTM attacks: all first run reduction algorithms on special lattices (``preproc'' time), then use the reduced vectors to either recover secrets (``recover'' time for ML/CC) or solve decision LWE (``decide'' time for MiTM). The ``preproc'' step is fully parallelizable, so the number of CPUs listed is the number of reduced lattices needed per attack. For example, the ML attack reduces $m \times n$ LWE matrices and needs $2$ million training samples (\S\ref{subsec:ml}). Each reduced matrix produces about $m+n$ samples\textemdash we ignore all-$0$ rows\textemdash so the ML attack requires roughly $2000000/(m+n)$ matrices and CPUs. The CC attack needs $100$K samples (but only $5K$ samples to solve Decision-LWE), so it only needs $100000/(m+n)$ matrices/CPUs. Table~\ref{tab:ml_cc_preproc} gives the average number of samples produced per reduced matrix for ML and CC. The MiTM attack needs $\tau$ short vectors, each obtained from a separate lattice, so it needs $\tau$ CPUs. 

For the ML and CC attacks, the ``recover'' step is parallelizable due to the cliff shifting approach described in \S\ref{subsec:ml}. For ML, we train separate models on datasets formed from the $n$ possible cliff shifts, using $n$ GPUs. For CC, we also run brute force on all $n$ shifted datasets, using $n$ GPUs. It is difficult to parallelize the table guessing step of MiTM due to memory constraints, so MiTM ``decide'' step runs on a single CPU. uSVP attacks are not parallelizable and do not have separate preprocess/recovery stages.

\vspace{-0.1cm}
\subsection{Analysis of Results}
\vspace{-0.1cm}

For all settings, the CC attack recovers secrets with the highest Hamming weights, slightly better than the ML attack, and using less compute. Unfortunately, the CC attack does not scale well to higher Hamming weights since it relies on exhaustive search to recover cruel bits.  Attack times (assuming full parallelization) are roughly equivalent for the ML and CC attacks, since the preprocessing time dominates for both approaches. Further improvements to the ML attack may allow it to scale to higher Hamming weights. 

The MiTM only solves the Decision-LWE approach, so it is not comparable without further work to convert to a Search-LWE algorithm. It also includes an exhaustive search subroutine which scales exponentially as the hamming weight grows. For the $h$'s it can decide, the MitM attack required the least compute\textemdash only $50$ CPUs, since $\tau=50$ short vectors are sufficient. However, all recovered MiTM secrets have $h' \le 6$. Even though $h' \le 8$ could work with our memory limits for \kyb{} settings (see Table~\ref{tab:both_mitm_mem}), the number of secret coordinate values in binomial secrets $(-2, -1, 0, 1, 2)$  that must be searched makes searches on $h'=8$ secrets take many days. 
The memory requirements for the MitM attack also scale badly as
the hamming weight increases.

In summary, { Cool\&Cruel is currently the best attack on our benchmark settings}.

\subsection{Actual vs. Estimated Performance} 
For the two attacks implemented in the lattice estimator (uSVP and dual hybrid MiTM), we also provide cost estimates from the estimator (commit \texttt{f18533a}) for each benchmark setting. We only present estimates for the Chen Nguyen cost model, which best approximates the \texttt{fplll} implementation of BKZ SVP and should most closely match our experimental results. Estimates can be found in Table~\ref{tab:kyber_he_estimates}. The Estimator natively supports sparse ternary $\ternsech$, but we add a function in \texttt{nd.py} to estimate attack performance on fixed $h$ binomial secrets, $\binomsech$ and $\normsech$. See Appendix~\ref{appx:functions} for details. A script to generate these estimates will be included in our open-source codebase. 

The Estimator predicts the uSVP attack should not be feasible (times are in years). Despite these predictions, we ran numerous uSVP experiments with much smaller-than-predicted blocksizes to see if they would work. None succeeded, despite running for over $2$ months on our compute cluster. However, we observed several interesting discrepancies between estimated and actual BKZ performance in these experiments, which are discussed in \S\ref{subsec:bkz} and Appendix~\ref{sec:BKZ2.0_timings}. 

The Estimator results given for the Dual Hybrid MiTM attack do not map particularly well to our real-world results. The Estimator under-predicts the time required to run one attack (e.g. 0.9 hours predicted vs. 65 actual for $n=1024$, $\log_2 q=26$, assuming full parallelization), but overestimates the number of repeats needed for high probability of attack success. Our attacks mostly succeed on the first try. If one multiplies concrete attack time by number of required machines, then Estimator time predictions are even more off\textemdash 0.9 hours vs. $65 \cdot 50 = 3250$ hrs = $135$ days for $n=1024$, $\log_2q = 26$. Furthermore, the Estimator-predicted $\tau$ and $\zeta$ values did not work well in practice\textemdash we only needed $\tau=50$ vectors to succeed, but the estimated $\zeta$ values were too small, resulting in very long scaled dual lattice reduction time. Additional engineering improvements to the MiTM attack may improve attack time costs, but future work should consider whether formulae for $\zeta$ and $\tau$ are accurate.

\section{Lessons Learned}
\label{sec:lessons}

Although the benchmark results of the prior section are valuable on their own, all our experimental work generating them yielded interesting insights about attack behavior. Here, we highlight a few of these experimental observations that we believe may be valuable to the research community. Future efforts to implement and evaluate other LWE attacks will likely yield fruitful observations to drive future research. 

\subsection{Q-ary Lattice Reduction}
\ejw{The Cool \& Cruel attack paper~\cite{nolte2024cool} showed that when reducing a lattice basis of the form:
\[
\mathbf{\Lambda} =
 \begin{pmatrix}
0 & q\cdot \mathbf{I}_n \\
\omega\cdot\mathbf{I}_m & \mathbf{A} 
\end{pmatrix}
\label{eq:Lambda_mat}
\] the short vectors are not always balanced, contrary to the common assumption in the literature. 
In particular, letting 
%\[
$\mathbf{ U\Lambda} = (\omega \mathbf R, \mathbf{RA} \mod q)$ 
%\] 
be the reduced matrix obtained by applying BKZ2.0 to
$\mathbf \Lambda$,
Figure \ref{fig:bkz_std} illustrates the standard deviations of the columns of $U\Lambda$ for $n=m=256$, $q=3329$, and $\omega=1$. It reveals that BKZ2.0 reduces vectors from right to left. Depending on the block size $\beta$, it terminates at a certain point, leaving a portion of the vector unreduced. ~\cite{nolte2024cool} used this to mount a powerful attack on sparse secrets.}

\begin{figure}[h]
    \centering
    \includegraphics[width=0.5\textwidth]{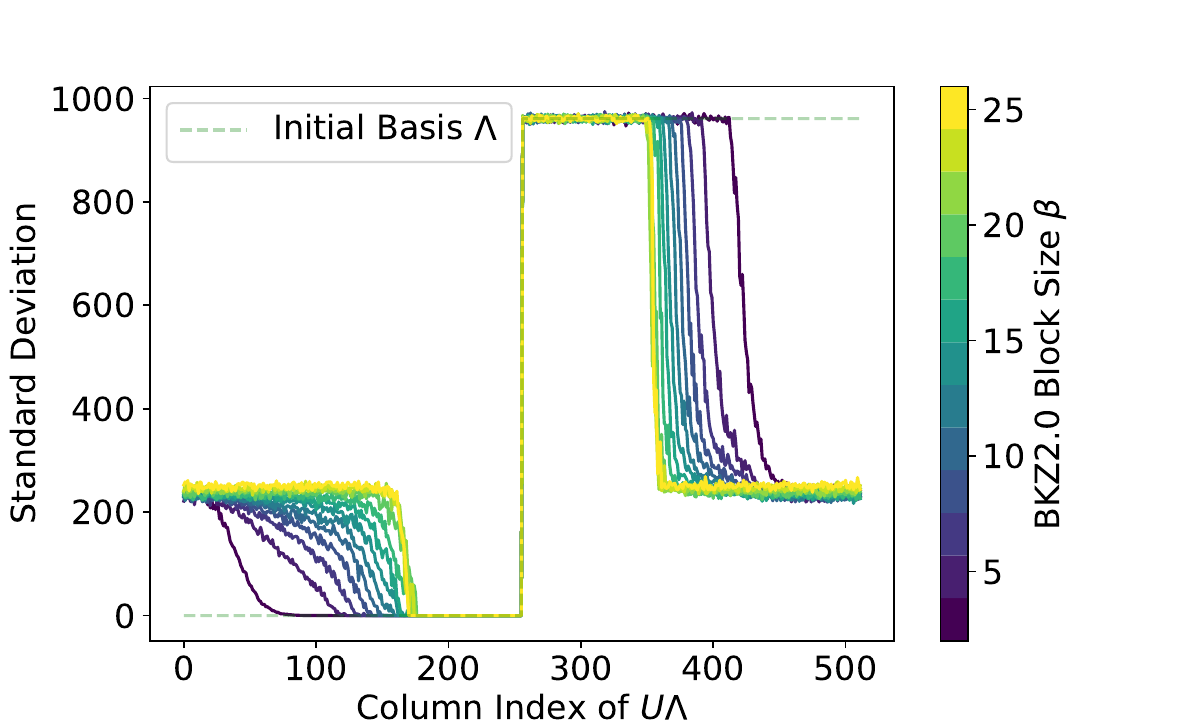}
    \vspace{-0.4cm}
    \caption{
     {\bf  Larger BKZ2.0 block size produces data with a smaller cliff.} As block size increases (color tends toward yellow), BKZ2.0 yields rows of $U\Lambda$ with more entries tending towards the reduced value. Figure shows parameters $n=m=256$, $q=3329$, $\omega=1$, and varying block size $\beta$.
    }
    \label{fig:bkz_std}
\end{figure}

\ejw{We make two observations extending~\cite{nolte2024cool}. First, we observe experimentally that as the block size $\beta$ increases, BKZ2.0 tends towards reducing all components of the initial basis, producing balanced vectors, see \cref{fig:bkz_std}.  }
Furthermore, \ejw{we see that BKZ2.0 underutilizes the later rows of the matrix $\mathbf{A}$.} 
The left half of \cref{fig:bkz_std} reveals that the rows of $\bf R$ are not balanced either. This side of the graph plots $\omega \mathbf R$, and we see that the last columns of $\mathbf R$ have only small values, \ejw{indicating that BKZ2.0 has not fully used the later rows of $\mathbf{A}$.} Future work should study this phenomenon and ways to exploit it.

\subsection{Discrepancy in BKZ2.0 timings}
\label{subsec:bkz}

Next, we highlight an interesting experimentally observed discrepancy in the predicted and actual BKZ2.0 loop time. We ran BKZ2.0 with $\beta$ ranging from $40$ to $54$ on q-ary-embedded (\cref{eq:qary}) ($m \times n$) LWE matrices with $n=512$, $m=712$, $q=3329$, and $\omega=4$ and measured the time it took for BKZ2.0 to complete one loop.  We then computed the predicted loop cycle/time for BKZ2.0 with these settings with two enumeration SVP cost models: CheNgu12~\cite{CN11_BKZ} and ABLR21~\cite{albrecht2021lattice} (see Appendix~\ref{sec:BKZ2.0_timings} for details). Concrete BKZ2.0 loop times and estimated timings are in Table~\ref{tab:concrete_bkz_512_12}.
\begin{table*}[t]
\resizebox{0.99\textwidth}{!}{
\begin{tabular}{lllllllllllll}
\toprule
BKZ $\beta$ & 30 & 40 & 45 & 46 & 47 & 48 & 49 & 50 & 51 & 52 & 53 & 54 \\ \midrule
predicted cycles~\cite{CN11_BKZ} & $2^{39.73}$ & $2^{39.77}$ & $2^{39.83}$ & $2^{39.85}$ & $2^{39.88}$ & $2^{39.92}$ & $2^{39.96}$ & $2^{40.01}$ & $2^{40.07}$ & $2^{40.14}$ & $2^{40.23}$ & $2^{40.34}$ \\
predicted time~\cite{CN11_BKZ} & 0.12 & 0.12 & 0.13 & 0.13 & 0.13 & 0.14 & 0.14 & 0.15 & 0.15 & 0.16 & 0.17 & 0.18 \\ \midrule
predicted cycles \cite{albrecht2021lattice} & $2^{42.95}$ & $2^{44.16}$ & $2^{45.05}$ & $2^{45.24}$ & $2^{45.45}$ & $2^{45.65}$ & $2^{45.87}$ & $2^{46.09}$ & $2^{46.32}$ & $2^{46.55}$ & $2^{46.79}$ & $2^{47.03}$ \\
predicted hours \cite{albrecht2021lattice} & 1.05 & 2.43 & 4.48 & 5.13 & 5.90 & 6.82 & 7.92 & 9.22 & 10.79 & 12.67 & 14.94 & 17.67 \\ \midrule
{\bf Actual (1st BKZ loop hrs)} & 0.53 & 0.78 & 1.73 & 1.55 & 2.6 & 5.3 & 9.2 & 21.8 & 38.2 & 67.4 & $>72$ & $>72$ \\ \bottomrule
\end{tabular} 
}
\caption{{\bf Concrete BKZ2 timings for $n=256$, $k=2$, $\log_2 q=12$.} To achieve practical speedup, three loops of \texttt{flatter} are run before switching to BKZ2, so reported BKZ2 times are for partially-reduced matrices. For $\beta > 52$, BKZ2 ran for $3$ days before hitting our computer cluster's time limit.}
\label{tab:concrete_bkz_512_12}
\end{table*}

Our experimental results do not closely match either cost model. We observe a sharp increase in the BKZ2.0 loop time starting around $\beta=49$. 
For $\beta \ge 52$, BKZ2.0 takes several {\em days} to run, sometimes failing to terminate within our $72$ hour cluster time limit. 

\ejw{This discrepancy may be due to some implementation issue or to problems with the estimates themselves. We observe that the cost model derived from~\cite{CN11_BKZ} is a linear curve fit to experimental results of BKZ2.0 runs on $n \le 250$. (see \texttt{CheNgu12} function in \texttt{reduction.py} of the lattice estimator). Other cost models, such as that of~\cite{albrecht2021lattice}, build on this curve fit. If the~\cite{CN11_BKZ} cost model does not generalize well beyond $n=250$, this could explain the error.} More rigorous experimentation with BKZ2.0 in higher dimensions and block size $\beta$ is needed to understand this.

\subsection{ML attacks recover secrets with $\le 3$ cruel bits}
\label{subsec:irwin_hall}

Although the Cool\&Cruel and ML attacks recover similar $h$ values, the CC attack's performance can be readily explained by a brute force scaling law. The ML attack's limitations are more mysterious. Since the ML attack trains on data reduced in the same manner as the Cool\&Cruel attack, we consider whether some property of the ``cruel'' region of reduced data affects secret recovery. Analyzing secrets through this lens, we find that ML models only ever recover secrets with $\le 3$ cruel bits. This pattern persists across secret distributions.

\begin{figure}[h]
    \centering
    \includegraphics[width=0.4\textwidth]{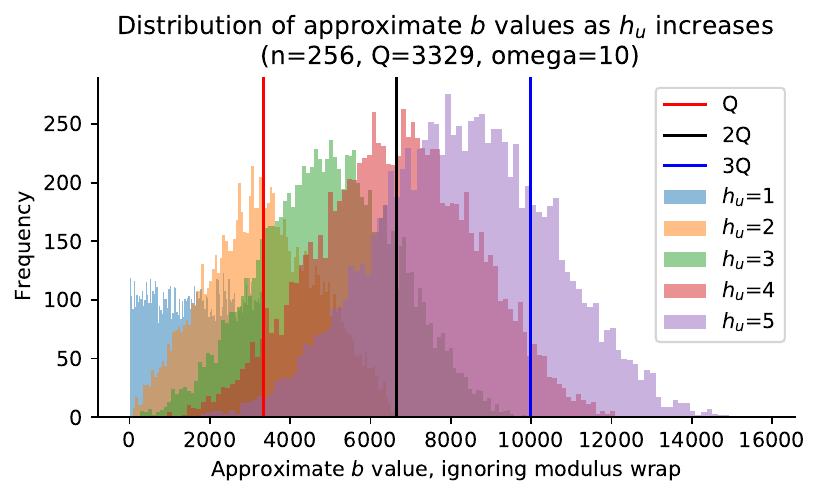}
    \includegraphics[width=0.4\textwidth]{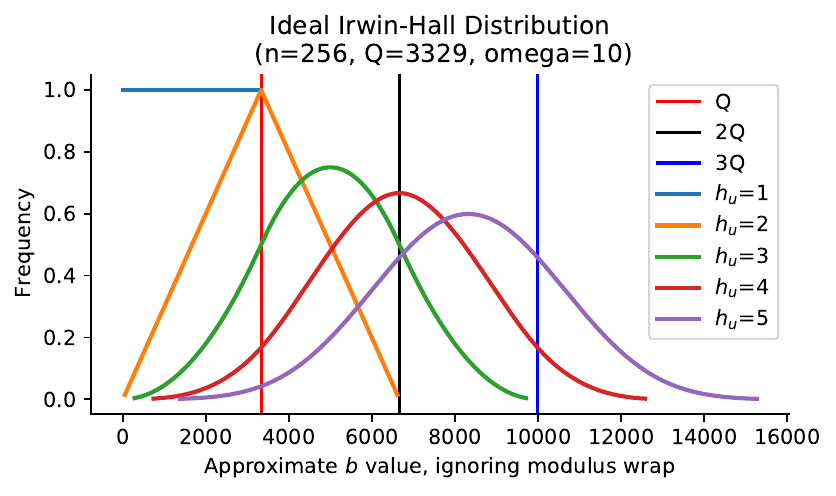}
    \vspace{-0.2cm}
    \caption{{\bf Visualization of empirical and ideal Irwin-Hall distribution for $n=256$, $k=1$, $\log_2q=12$ setting.}}
    \label{fig:irwin_hall}
    \vspace{-0.2cm}
\end{figure}
We believe this phenomenon can be explained by distribution of sums of uniform random elements $\mod q$. Modular addition of uniform random elements in $\mathbb{Z}_q$ is well-described by a modified Irwin-Hall distribution, which describes the distribution of sums of $n$ uniform random elements $U(0,1)$: $ X = \sum_{k=1}^n U_k$. Multiplying by $q$ yields a distribution describing sums of $n$ uniform random elements mod $q$. Figure~\ref{fig:irwin_hall} visualizes this for a $n=256$, $\log_2 q=12$ dataset. From this graph, we observe that sums of $3$ random uniform variables usually fall in range $[q, 2q]$. Since cool region elements hardly affect this sum, this means that secrets with $3$ cruel bits produce mostly $b$ values that only wrap once around the modulus. For ML models, this means little or no modular arithmetic must be learned.

This explains why the ML attack only recovers $3$ cruel bit secrets: models struggle to learn modular arithmetic (as first observed in~\cite{verde}), and secrets with $> 3$ cruel bits require models to understand that certain $b$ involve sums which ``wrap around'' the modulus. Prior work has observed models' poor performance on modular arithmetic setups~\cite{jelassi2023length, palamasinvestigating, gromov2023grokking, lauter2024machine}, and this result confirms that the difficulty persists. Future improvements to the ML attack could focus on strategies to help models learn modular arithmetic.

\subsection{Effect of bad PRNGs on attack performance}
\label{subsec:badrng}

Finally, we note that bad pseudo-random number generators (PRNGs) can make LWE secrets easier to recover.  We observed experimentally that \texttt{flatter} reduction performed {\em significantly} better on LWE matrices $\bf A$ $\in \mathbb{Z}^{m \times n}_q$ generated by the C \texttt{random} library than on otherwise-identical $\bf A$ matrices generated by numpy's \texttt{random} library (see Figure~\ref{fig:badrng_n128}). Furthermore, ML models trained on \texttt{flatter}-reduced, $\texttt{C random}$ $\bf A$ matrices recover binomial secrets with $h \le 90$, see Table~\ref{tab:binom}, a feat not possible for the numpy $\texttt{random}$ $\bf A$ matrices.  In both cases, $\bf A$ matrices are generated row-by-row, filling the $n$ slots of row $1$ with random integers mod $q$, then filling row $2$, etc.

\begin{table}[h]
    \centering
    \begin{tabular}{cccc}
    \toprule
        $h$ & 50 & 70 & 90\\ \midrule
        attack time (hrs) & 3 & 3.75 & 3 \\
        recovery rate & 5/5 & 5/5 & 5/5 \\ \bottomrule
    \end{tabular}
    \caption{{\bf ML attack recovers binomial secrets with up to $h=90$ for $n=256$, $k=1$, $\log_2 q=12$ data when LWE data is generated with the \texttt{C random} LCG.} $h$ = Hamming weight of recovered secret, Time = avg hours to secret recovery (excluding preprocessing), Recovery rate = Secrets recovered / attempted.}
    \label{tab:binom}
\end{table}

\begin{figure}[h]
    \centering
    \includegraphics[width=0.35\textwidth]{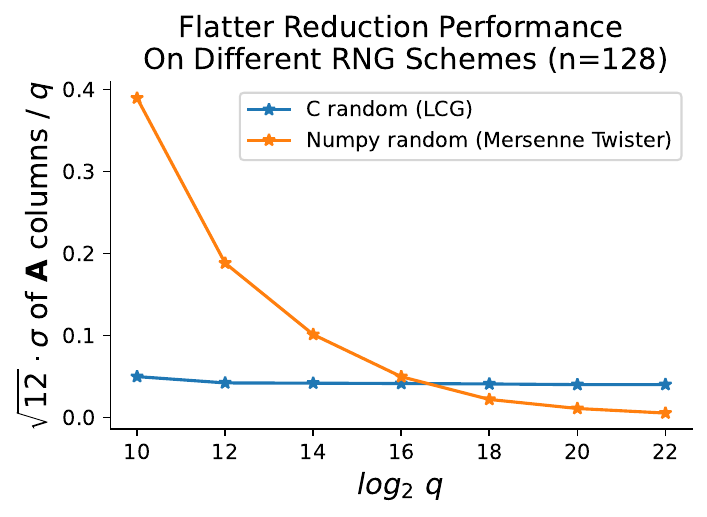}
    \vspace{-0.1cm}
    \caption{{\bf The \texttt{flatter} algorithm performs significantly better on LWE matrices generated with the \texttt{C random} LCG than with the \texttt{numpy random} Mersenne twister.} Y-axis shows reduction in standard deviation of ${\bf A}$ elements vs. uniform random standard deviation (lower values indicate stronger reduction).}  
    \label{fig:badrng_n128}
\end{figure}

After observing this performance discrepancy, we found that the \texttt{C} {\tt random()} function is implemented via the BSD linear-congruential generator (LCG) by many computer distributions, including MacOS Clang used by {\tt gcc} under Xcode and GNU {\tt gcc} under Linux, while \texttt{numpy} random uses a Mersenne Twister. Prior work showed that LCGs produce predictable outputs~\cite{marsaglia1968random}, and consequently should not be used in cryptographic settings. However, no prior work has observed this specific vulnerability of LCG-generated LWE matrices to lattice reduction attacks. 

An LCG algorithm is defined by the recurrence relation: $x_{i+1} = x_i a +c$ mod $m$, where the modulus $m$, the multiplier $a$ and the increment $c$ are non-negative integer constants. It can be shown via an induction argument that a matrix $\bf A$ $\in \mathbb{Z}^{m \times n}_q$ generated row by row via an LCG, as described, has columns also generated by a related LCG, since $x_{i+n} = x_i  a^n + (a^{n-1} + a^{n-2} + \dots +1)c$ mod $m$. Furthermore, if we view LCG generated vectors as points in the cube in the corresponding dimension, as described in~\cite{marsaglia1968random}, then the points will lie on equally distanced parallel hyper-planes of the form $c_1 p_1 + c_2 p_2 + \dots + c_l p_l = 0, \pm m, \pm 2m, \dots$. The number of these hyper-planes, with the parameters of the LCG used by \texttt{C random}, and LWE dimensions our benchmarks consider, is small (two digits).The distance between them can also be shown to be small compared to norms of (row or column) vectors from $\bf A$. Thus, the matrix has many structures that potentially could be exploited by lattice reduction.

We highlight this issue because even though PRNGs are well-known to be bad for cryptography, this advice might not always be followed.  Furthermore, the fact that LCGs are included as the standard RNG in some libraries increases the likelihood that they may be accidentally used. Using LCG-generated $\bf A$ matrices in LWE attack development would make the attack appear exceptionally good, while LCG use in LWE encryption schemes would create significant vulnerabilities. We observe at least one case of a LCG PRNG used in real-world LWE crypto: the \texttt{C++} \texttt{rand} function is used in the testing functions of the HEEAN library~\cite{params_heean}\footnote{github.com/snucrypto/HEAAN/blob/master/HEAAN/src/EvaluatorUtils.cpp}. While this use poses no imminent danger, we highlight this as a cautionary tale for implementers of Kyber and HE. 

 \section{Join our LWE Attack Benchmarking Effort}
\label{sec:codebase}

To accompany the benchmark settings and evaluations in this paper, we provide an open source codebase implementing the attacks we evaluate. We hope that by making our code available to the public, others will join us in establishing experimental benchmarks for LWE attacks. Our code can be found at \url{https://github.com/facebookresearch/LWE-benchmarking} and an associated website is at \url{https://facebookresearch.github.io/LWE-benchmarking/}.

\para{Codebase Overview.} Our codebase contains (1) code to preprocess and generate LWE, RLWE, and MLWE data and (2) implementations of four different attacks: transformer-based ML attack, dual hybrid MiTM attack, USVP attack, and Cruel and Cool (CC) attack. To run these attacks, a user would first prepare the data by running the preprocessing step and generating LWE ($\bf A$, $\bf b$) pairs and associated secrets. Next, a user can run any of the four attacks on the generated data by running that attack's script with the data path and relevant parameters.
More details on how to set up and run the code are provided in the README.

\para{Contributing.} We invite contributors to reproduce our results, improve on these methods, and/or implement new LWE attacks. We actively welcome pull requests with new or improved attacks or code improvements. 
Please include the code and instructions on how to reproduce the results in the pull request.
Please document added code, provide proof of testing, and follow a style similar to the rest of the repository. We will also use Github issues to track public bugs. Please provide a clear description of the bug and instructions on how to reproduce the problem. See the CONTRIBUTING file in our codebase for instructions on how to contribute.

We will also maintain a centralized leaderboard of the best performing attacks on LWE along the axes of time, space, and compute.  We will update the leaderboard on our associated challenge website accordingly. 

\section{Conclusion and Future Work}
\label{sec:discussion}

This paper demonstrates the first successful LWE secret recovery on standardized \texttt{KYBER} and HE parameters--not yet general secrets but small, sparse secrets. For example, in the setting $n=256$, $k=2$, $\log_2q = 12$ we recover binomial secrets with Hamming weight $h \le 11$ in $<36$ hours (parallelized compute); for the HE setting $n=1024$, $\log_2 q=29$, we recover Hamming weight $h=9$ secrets in $13$ hours. This paper provides the first benchmarks of LWE attack performance on near-real-world settings. 

This paper also makes meaningful contributions through its efforts implementing and scaling up the attacks evaluated, yielding valuable lessons learned that can inform future research. We hope the insights shared from our work implementing these attacks will aid and inspire other researchers in the lattice community to join us in this benchmarking endeavour. Topics for future work include:
% and 

\para{Optimizing attack implementations.} Although we do our best to fairly compare the four attacks we evaluate, there are inevitable inefficiencies in our implementations. We expect that additional engineering would make these attacks more efficient. For example, speeding up the enumeration SVP in BKZ2.0 (via a GPU implementation like~\cite{pohmannfpll}) would greatly improve the lattice reduction time for all attacks. 
\para{Revisiting theoretical estimates with experimental insights.} In several instances, theoretical predictions do not match experimental observations. For example, the Estimator over-estimates the number of short vectors needed but under-estimates time needed for successful Dual Hybrid MiTM attacks. We also observe discrepancies between predicted and actual BKZ2.0 times, as shown in \S\ref{subsec:bkz} and Appendix~\ref{sec:BKZ2.0_timings}. \ejw{Rigorous work is needed to better understand, and eventually correct, these discrepancies. In particular, future work should ensure theoretical assumptions align with real-world behavior.} 

\para{Implement additional attacks in open source codebase.} This work evaluates a subset of relevant attacks on LWE, and important future work involves implementing additional attacks for evaluation. Everyone is welcome to contribute their own attack implementation to the open source codebase we release with this paper. Interesting attacks to consider implementing include, but are not limited to, Bounded Distance Decoding (BDD) attacks, primal hybrid attacks, and the dual hybrid approach of~\cite{bilu2021} that uses matrix multiplication and pruning instead of MiTM for guessing. 

\para{Use benchmark settings to evaluate new attacks.} The benchmark settings that we propose are not merely retrospective, allowing comparison of already-existent LWE attacks. Rather, they should enable more robust understanding how new attacks fit into the research landscape. We encourage the research community to adopt the benchmark settings proposed here as part of a standard evaluation set for all new attacks, alongside theoretical estimates.

\vspace{-0.2cm}
\subsection*{Acknowledgements}
\vspace{-0.2cm}
\noindent We thank Francois Charton, Niklas Nolte, and Mark Tygert for suggestions and contributions.


\newpage

\bibliographystyle{plain}
\bibliography{references}

% \newpage

% \section{Appendix}
\appendix
\label{sec:appendix}
\def\thesubsectiondis{\Alph{subsection}.} % B.
\renewcommand\thesubsectiondis{\Alph{subsection}.}
\subsection{Modifications to Lattice Estimator}
\label{appx:functions}

As of commit \texttt{00ec72ce}, the Lattice Estimator only supports sparse ternary secrets (through the \texttt{SparseTernary} function in \texttt{nd.py}). To estimate attack performance on benchmarks proposed in this attack, we add a \texttt{SparseCenteredBinomial} function to \texttt{nd.py}, which enable estimation on sparse binomial and Gaussian secrets. Code for these is available upon request. 

\subsection{Concrete BKZ2.0 Timings}
\label{sec:BKZ2.0_timings}

Although we found that uSVP attacks, as predicted, did not succeed in reasonable time for our benchmark settings, we did make some interesting observations from these experiments. Primarily, we noticed that the lattice estimator can underestimate time needed for enumeration-based BKZ2.0 lattice reduction, even when incorporating \texttt{flatter} to speed up BKZ2.0, for $\log_2 q > 40$.

We compare concrete \texttt{fplll} BKZ2.0 timings to estimates times for enumeration-based BKZ2.0. We used two models for estimation: the Chen-Nguyen (CheNgu) model~\cite{CN11_BKZ}, and the ALBR21 model~\cite{albrecht2021lattice}. The CheNgu model is based on a curve fit to concrete results in dimension $n\le 250$ provided in~\cite{CN11_BKZ}, and is likely the closest estimate of the actual enumeration model used in \texttt{fplll} BKZ2.0. According to this model, one loop of BKZ2.0 will visit $2^{(0.18728\beta \cdot \log_2(\beta) - 1.019\cdot \beta + 16.10)}$ enumeration nodes, where $\beta$ is BKZ2.0 block size. To get a concrete time estimate, we must multiply by the cost of visiting each node. According to the header comment in the \texttt{CheNgu} function in \texttt{reduction.py} in the lattice estimator, this cost is 64. 
The Estimator multiplies this by a “repeats” factor of $8n$, which is the number of estimated SVP calls within one loop of BKZ2.0-Beta. According to a comment in the Estimator, this is loosely based on results from Yuanmi Chen’s 2013 PhD thesis~\cite{chen2013reduction}. To this, we add the cost of LLL, which is $n^3 (\log q)^2$, 
where $n$ is lattice dimension. 

For completeness, we also include the ABLR21 cost model~\cite{albrecht2021lattice}, which proposes faster enumeration strategies, improving upon the asymptotic cost of Chen-Nguyen. The following cost model is derived from simulations in this paper. If $\beta \le 97$ or $1.5\beta \ge n$, the cost is $64 * 2^{(0.1839\beta \cdot \log_2(\beta) - 1.077 \cdot \beta + 29.12)}$ (this includes the factor of $64$ for node visitation), otherwise it is $64 * 2^{(0.125\beta \cdot \log_2(\beta) - 0.654 \cdot \beta + 25.84)}$. Since \texttt{fplll} does not use the enumeration strategy from this paper, we expect this cost estimate will underestimate concrete \texttt{fplll} performance, but we include it for completeness, and multiply it by the cost of LLL and expected repeats as above.  

We use these to derive time estimates for each BKZ2.0 loop for all the $n$, $q$, and $\beta$ values for which we ran concrete uSVP experiments. As previously, we convert ROP cycles to time by dividing out the cycle speed of our machines ($2.1GHz$). Since we reduce lattices with Kannan's embedding, the effective lattice dimension is $m+n+1$, where $m=0.875n$ = the number of LWE samples per embedded lattice. We set \texttt{BKZ\_MAX\_TIME = 60}, which means that unless the algorithm takes $< 60$ seconds, it will return after each loop. This ensures that we can accurately time each BKZ2.0 loop, as well as the time to uSVP solution (although this only occurs for small matrices).

\begin{table}[h]
\centering
\resizebox{0.5\textwidth}{!}{
\begin{tabular}{llll} 
\toprule
$n$ & 64 & 128 & 256 \\
$q$ & 967 & 11197 & 397921 \\
BKZ $\beta$ & 30 & 30 & 50 \\ \midrule
predicted cycles~\cite{CN11_BKZ} & $2^{31.3}$ & $2^{33.9}$ & $2^{37.6}$ \\
predicted time~\cite{CN11_BKZ} & 1.2 s & 7.5 s & 1.6 mins \\ \midrule
predicted cycles~\cite{albrecht2021lattice} & $2^{39.8}$ & $2^{40.8}$ & $2^{45.1}$ \\
predicted time~\cite{albrecht2021lattice} & 7 mins & 14 mins & 4.6 hrs \\ \midrule
actual time, first BKZ2.0 loop & $\sim$30 s & 1 minute & 3.3 hrs \\ 
secret found? & Yes (1 loop) & Yes (1 loop) & No \\ \bottomrule
\end{tabular}
}
\caption{{\bf Estimated vs. actual times for first loop of BKZ2.0} $n \le 256$. We convert predicted cycles to concrete times by dividing by the cycle speed of our CPUs (2.1 GHz), following~\cite{albrect_24_talk}. Since BKZ2.0 runs on uSVP problems, we also report whether the secret is recovered.}
\label{tab:small_concrete_bkz}
\end{table}

\begin{table*}[h]
\resizebox{0.99\textwidth}{!}{
\begin{tabular}{lcccccc@{\hskip 0.25cm}|@{\hskip 0.25cm}cccccc@{\hskip 0.25cm}|@{\hskip 0.25cm}cc}
\toprule
%n & \multicolumn{6}{c}{512} & \multicolumn{6}{c}{768} & \multicolumn{2}{c}{1024} \\[0.2cm] %\cmidrule{2-15}
($n$, $\log_2 q$) & \multicolumn{2}{c}{(512, 32)}  & \multicolumn{2}{c}{(512, 34)} & \multicolumn{2}{c}{\hspace{-0.35cm}(512, 41)} & \multicolumn{2}{c}{(768, 45)} & \multicolumn{2}{c}{(768, 50)} & \multicolumn{2}{c}{(768, 53)} & (1024, 34) & (1024, 50) \\\midrule
blocksize $\beta$ & 70 & 74 & 60 & 64 & 40 & 45  & 70 & 80 & 60 & 68 & 50 & 58  & 70 & 75 \\ \midrule
predicted cycles~\cite{CN11_BKZ} & $2^{44.8}$ & $2^{46.3}$ & $2^{42.4}$ & $2^{43.1}$ & $2^{41.7}$  & $2^{41.7}$ & $2^{45.6}$ & $2^{49.4}$ & $2^{43.7}$ & $2^{45.0 }$& $2^{43.5}$ &$ 2^{43.7} $& $2^{46.3}$ &$ 2^{48.7}$ \\ 
predicted hrs~\cite{CN11_BKZ} & 4.2  & 11.8  & 0.74  & 1.21  & 0.5  & 0.5  & 7.1  & 101.2  & 2.0  & 4.8  & 1.7  & 1.9  & 11.2  & 34.1  \\ \midrule
predicted cycles~\cite{albrecht2021lattice} & $2^{51.5}$ & $2^{52.8}$ & $2^{48.6}$ & $2^{49.7}$ & $2^{44.3}$ & $2^{45.2}$  & $2^{52.1}$ & $2^{55.5}$ & $2^{49.2} $& $2^{51.5}$ & $2^{46.8.5 }$ & $2^{48.7}$ & $2^{52.5}$ & $2^{54.2}$ \\
predicted hours~\cite{albrecht2021lattice} & 404  & 987  & 52  & 114  & 2.8  & 4.8  & 606  & 6107  & 79.4 & 394  & 15.2  & 55.1  & 809  & 2488  \\ \midrule
First BKZ loop (hrs) & $370$  & $1083$ & $46.2$  & $109$  & $46.6$  & $47.3$  & $1195$  & $>1512$  & $227$  & $1087$  & $219 $ &$ 236 $ &$ >1512 $ & $>1512$ \\ \bottomrule
\end{tabular}
}
\caption{{\bf Estimated vs. actual times for first loop of BKZ2.0 ($n \ge 512$)}. Subsequent BKZ2.0 loops often take much less time. For experiments that have not completed a single BKZ2.0 loop, we write $> x$, where $x$ is the number of hours elapsed before the day of manuscript submission.}
\label{tab:big_concrete_bkz}
\end{table*}

\begin{table*}[]
\centering
\begin{tabular}{lllllll}
\toprule
($n$, $\log_2 q$)  & (768, 35) & (1024, 26) & (1024, 29) & (1024, 34) & (1024, 45) & (1024, 50) \\ \midrule
%BKZ $\beta$ & 18 & 18 & 18 & 18 & 18 & 18 \\ \midrule
predicted cycles~\cite{CN11_BKZ} & $2^{41.7}$ & $2^{42.1}$ & $2^{42.4}$ & $2^{42.9}$ & $2^{43.7}$ & $2^{44.0}$ \\
predicted time~\cite{CN11_BKZ} & 0.5 hrs & 0.6 hrs & 0.7 hrs & 1.0 hrs & 1.8 hrs & 2.2 hrs \\ \midrule
predicted cycles~\cite{albrecht2021lattice} & $2^{43.5}$ & $2^{43.9}$ & $2^{44.0}$ & $2^{44.2}$ & $2^{44.6}$ & $2^{44.8}$ \\
predicted time~\cite{albrecht2021lattice} & 1.6 hrs & 2.1 hrs & 2.2 hrs & 2.5 hrs & 3.2 hrs & 3.7 hrs \\ \midrule
First loop flatter & 12.4 hrs & 29.7 hrs & 30.8 hrs & 17.5 hrs & 17.3 hrs & 23.8 hrs \\
First loop BKZ & - & 41.6 hrs & 34.1 hrs & 8.1 hrs & 2.9 hrs & - \\ \bottomrule
\end{tabular}
\caption{{\bf Estimated vs. actual times for first loop of BKZ2.0 with $\beta=18$ for large $n, \log_2q$ parameter settings, including those used in benchmark evaluation.} All experiments are run with BKZ2.0 with $\beta=18$. To achieve practical speedup, three loops of \texttt{flatter} are run before running the first loop of BKZ2.0, so reported BKZ2.0 times are for partially-reduced matrices. `-' indicates using flatter alone allowed us to reach the target reduction level, so reduction terminated before BKZ2.0 is run.}
\label{tab:datasets_bkz}
\end{table*}

Tables~\ref{tab:small_concrete_bkz} and~\ref{tab:big_concrete_bkz} compare concrete performance times to predicted times for small $n \le 256$ and large $n \ge 512$, respectively. For both small $n$ and $q$, estimates are fairly accurate. However, for $\log_2 q > 45$, estimates consistently underpredict BKZ2.0 time.  Finally, in Table~\ref{tab:datasets_bkz} we report predicted vs actual reduction times for the $n>512$ parameter settings, including some used in our benchmarks. For these, since blocksize $\beta$ is small ($\beta=18$), the dominant cost is that of LLL, and the LLL cost in the Estimator is determined by $B = \log_2q$. Thus, there is a slight increase in estimated time as $\log_2 q$ increases, but the estimates still under-predict the required times observed in practice. 

These discrepancies indicate a need for more in-depth consideration of the role of $\log_2q$ in the timing of BKZ2.0. Theoretical models for BKZ2.0 timing do not consider integer bitsize, although estimates of LLL time do.

\subsection{Cliff Shifting and Cliff Splitting}\label{appx:cliffshift}

The Cool and Cruel distinguishing attack \cite{nolte2024cool} exploits the algebraic structure of the 2-power cyclotomic Ring-LWE to strategically ``shift" the experimentally observed ``cliff'' in preprocessed LWE data to find an optimal window with low Hamming weight. Each index of a skew-circulant matrix formed from a reduced LWE sample has the cliff appear in a different sliding window. By searching through circulant indices $[1, \ldots n]$ and finding one with low Hamming weight in the cliff region, the brute force part of the attack can be made faster. In practice, this requires running a brute force attack on datasets composed of elements at each circulant index until a low Hamming weight region is found, increasing attack cost. 
Here, we provide more details on cliff shifting.

\subsubsection*{Cliff shifting: free short(ish) vectors}
Given a set of Ring-LWE samples $(a^{(i)}(x), b^{(i)}(x))_i$ where $a^{(i)}(x) \in R_q = \mathbb{Z}_q[X]/(X^n+1)$, and using the embedding $\mathbf a = (a_0, a_1, \dots, a_{n-1})$, a lattice reduction algorithm is applied to an embedded version of the lattice spanned by $(\mathbf a^{(1)}, \mathbf a^{(2)}, \dots, \mathbf a^{(m)})$ also noted as $\mathbf A \in \mathbb{Z}_q^{m\times n}$ where row $i$ corresponds to $\mathbf a^{(i)}$. Denote a short vector found by reducing the embedded matrix $\Lambda$ as $(\mathbf y', \mathbf y) \in \mathbb{Z}^{m+n}$ such that $\mathbf y=\mathbf y'\cdot \mathbf A \mod q$ ($\mathbf y'$ is a row of the $\mathbf R$ matrix defined in \cref{subsec:ml})
We can describe the cliff result of \cite{nolte2024cool} by defining $\mathbf y$ as having 2 components $(\mathbf y_u, \mathbf y_r)$ where $\mathbf y_r \in \mathbb{Z}_q^{n_r = n-n_u}$ is short while $\mathbf y_u \in \mathbb{Z}_q^{n_u}$ remains unreduced. $n_r$ and $n_u = n - n_r$ are the number of reduced and unreduced components in $\bf y$, respectively.

For $y(x) \in R_q$, (whose coefficients are $\mathbf y$), the $n-l^{th}$ line in the flipped Skew-circulant matrix can be described by operation $x^l$, where $x^l y(x) = \sum_{k=0}^{n-1} (-1)^{\floor{\frac{k+l}{n}}}y_{k} x^{k+l[n]}$. Note that the elements $x^l y(x)$ have the same L2 norm $||y(x)||_2$. For this reason, given the short vector $\mathbf y$ that represents the polynomial $y(x)$, we have $n$ short vectors $\mathbf y^{\rightarrow l} = (-\mathbf y_r^{[n_r-l:n_r)}, \mathbf y_u, \mathbf y_r^{[0: n_r-l)})$ for $0 \leq l \leq n-1$, where $\rightarrow$ denotes the Skew-circulant shifting operation $x^l$.

We denote by $\mathcal{D}$ the pre-processed dataset and $\mathcal{D}_{\rightarrow l}$ the same dataset shifted by $x^l$. Our modified version of the ML attack~\cite{stevens2024fresca} trains a model on each shifted dataset. Since at least one of the datasets is much easier than the others (see {\bf Sparse secret cruel bits} below), one model will likely recover the secret first, after which we terminate training. % of all models. 

\subsubsection*{Cliff Splitting} For Module-LWE, the same attack can be run using a technique we introduce called `cliff splitting'. This applies a permutation $P\in \mathbb{Z}^{kn\times kn}$ to the initial vectors before reduction, and then applying its inverse after reduction. The result is a set of short vectors where each module component exhibits a similar profile. We describe this technique below.

Let the $R_q$-module $\mathcal{M} = R_q^k$, and $\mathbf (\mathfrak a, b =\mathfrak a\cdot \mathfrak s + e)$ be a Module-LWE sample. For $\mathfrak a = (a_1(x), a_2(x), \dots, a_k(x)) \in \mathcal{M}$, we consider the coefficient embedding of each component in one large vector
$\mathbf a = (\mathbf a_{10}, \mathbf a_{11}, \dots, \mathbf a_{1n-1}, \dots, \mathbf a_{k1},\mathbf a_{k2}, \dots, \mathbf a_{kn-1}) \in \mathbb{Z}_q^{kn}$ and run preprocessing on ``LWE-like'' matrices in $\mathbb{Z}_q^{m \times kn}$ formed by sampling $m$ of these embedded vectors. After preprocessing, similarly to the Ring case, if $\mathfrak a  \in \mathcal{M}$ is short, then $x^l\mathfrak a = (x^la_1(x), x^la_2(x), \dots, x^la_k(x))$ whose embedding is $\mathbf a^{\rightarrow l} = (\mathbf a^{\rightarrow l}_1, \mathbf a^{\rightarrow l}_2, \dots, \mathbf a^{\rightarrow l}_k)$ is also short.

With these short vectors, we can now perform cliff splitting. Let $\nu = \frac{N_u}{k}$, where $N_u$ is the cliff size of reduced lattice in dimension $kn$. We assume that $N_u \mod k = 0$ for simplicity in notation. Given a $kn$-vector $\mathbf{a} = (\mathbf a_1, \mathbf a_2, \dots, \mathbf a_k)$, we can split each component $\mathbf a_i$ into two parts: $\mathbf a_i^{[0:\nu)}$ and $\mathbf a_i^{[\nu, n)}$. This yields:
\begin{align*}
\mathbf{a} &= (\mathbf a_1^{[0:\nu)},\mathbf a_1^{[\nu, n)}, \mathbf a_2^{[0:\nu)},\mathbf a_2^{[\nu, n)}, \dots, \mathbf a_k^{[0:\nu)},\mathbf a_k^{[\nu, n)})
\end{align*}
We then apply the permutation $P$ to $\mathbf{a}$, which rearranges the components of $\mathbf{a}$ such that all unreduced regions come first:
\begin{align*}
\mathbf{a}P &= (\mathbf a_1^{[0:\nu)}, \mathbf a_2^{[0:\nu)},\dots, \mathbf a_k^{[0:\nu)},\mathbf a_1^{[\nu, n)}, \mathbf a_2^{[\nu, n)},\dots, \mathbf a_k^{[\nu, n)})
\end{align*}
After applying the lattice reduction to the permuted vectors, we apply $P^{-1}$ to obtain the final dataset $\mathcal{D}$ and the shifted datasets $\mathcal{D}_{\rightarrow l}$ for the ML~\cite{stevens2024fresca} or CC \cite{nolte2024cool} attacks.

\subsubsection*{Sparse secret cruel bits}
\label{appx:h_s}
To assess the advantage of Module-LWE over LWE, we begin by defining the partial Hamming weight of the secret $\mathbf s = (\mathbf s_1, \mathbf s_2, \dots, \mathbf s_k)$ where $\mathbf s$ is the embedding of $\mathfrak s \in \mathcal{M}^\vee$. This is done by considering the same window in each module component, defined as:
%\begin{align}
$h_{\nu, w}(\mathbf s) := \sum_{i=1}^{k} \sum_{j=w}^{w+\nu-1} \mathds{1}_{\{\mathbf s_{ij[n]} != 0\}}$
%\label{eq:hui}
%\end{align}
for Module-LWE and 
%\begin{align}
$h_{\nu, w}(\mathbf s) = \sum_{j=w}^{w+\nu-1} \mathds{1}_{\{\mathbf s_{j[n]} != 0\}}$ for Ring-LWE.
%\label{eq:hui_he}
%\end{align}
We then define $h_{\nu}^*(\mathbf s)$ as the minimum partial Hamming weight over all windows, and $w^*$ as the window that minimizes the partial Hamming weight:
\begin{align*}
h_{\nu}^*(\mathbf s) = \min_{0\leq w < n} h_{\nu, w}(\mathbf s),\quad
w^* = \arg\min_{0\leq w < n} h_{\nu, w}(\mathbf s)
\end{align*}

Colloquially, the CC attack defines $h_{\nu, w}(\mathbf s)$ as the ``cruel bits'' of a secret, and seeks the window with the fewest cruel bits.  The attack is then carried out on all datasets, including $\mathcal{D}_{\rightarrow w^*}$ whose vectors $\mathbf a$ have the fewest cruel secret bits in their un-reduced entries. % have their un-reduced entries coincide with the sparsest part of the secret $\mathbf s$ (e.g. fewest cruel bits). 
\cite{nolte2024cool} applies brute force on secret windows of size $N_u=k\nu$ starting with Hamming weight 0 and increasing from here. Although the value of $h_{\nu}^*(\mathbf s)$ is unknown, the attack can be halted as soon as a secret with Hamming weight $h_{\nu}^*(\mathbf s)$ is found. This reduces the search space and increases attack efficiency.

Experimentally, we find that attacks only conclude in reasonable time when $h_{\nu}^*(\mathbf s) \le 3$ for the ML attack and $h_{\nu}^*(\mathbf s) \le 4$ for the CC attack. We estimate those probabilities in Tables \ref{tab:kyber_probs} and \ref{tab:he_probs}.

\begin{table}[t]
\resizebox{0.49\textwidth}{!}{
\begin{tabular}{lcc|cc|cc}
\toprule
%& \multicolumn{4}{c}{$n=256, k = 2$} & \multicolumn{2}{c}{$n=256, k = 3$}\\ %\cmidrule{2-7}
 & \multicolumn{2}{c}{($k=2$, $\log q = 12$)} & \multicolumn{2}{c}{($k=2$, $\log q = 28$)} & \multicolumn{2}{c}{($k=3$, $\log q = 35$)} \\
\cmidrule{2-7}

 & $h=9$ & $h=11$ & $h=18$ & $h=25$ & $h=16$ & $h=19$ \\ 
\midrule
3 cruel bits & 14.3 & 2.0 & 19.1 & 1.2 & 26.3 & 8.6 \\
4 cruel bits & 51.4 & 11.3 & 47.9 & 4.9 & 60.2 & 26.0 \\
5 cruel bits & 96.2 & 40.1 & 83.5 & 14.9 & 93.0 & 56.4 \\ 
\bottomrule
\end{tabular}
}
\caption{{\bf Percent chance that secrets with Hamming weight $h$ have $\le x$ cruel bits ($h_{\nu}^*(\mathbf s) $) for Kyber settings ($n=256$ for all).} 
These represent the success probabilities of the CC/AI attacks given a compute budget measured in $x$. For an MLWE instance with $k=2, \log Q=28$ and a secret with $h=25$, if we run the brute force attack on all secret candidates with up $x = 5$ cruel bits, the attack would succeed with $15\% $ probability.}

\label{tab:kyber_probs}
\end{table}

\begin{table}[t]
\resizebox{0.49\textwidth}{!}{
\centering
\begin{tabular}{lcc|cc|cc}
\toprule
%& \multicolumn{6}{c}{$n = 1024$} \\
 & \multicolumn{2}{c}{$\log q = 26$} & \multicolumn{2}{c}{$\log q = 29$} & \multicolumn{2}{c}{$\log q = 50$} \\
\midrule
 & $h=8$ & $h=12$ & $h=10$ & $h=12$ & $h=17$ & $h=20$ \\ 
\midrule
3 cruel bits & 43.4 & 1.7  & 16.8 & 3.6 & 14.8 & 4.3\\
4 cruel bits & 93.1 & 8.9 & 52.5 & 16.0 & 39.9 & 14.5\\
5 cruel bits & 100.0 & 30.1 & 94.7 & 46.8 & 75.2 & 36.8 \\
\bottomrule
\end{tabular}
}
\caption{ {\bf Percent chance that secrets with Hamming weight $h$ have $\le x$ cruel bits ($h_{\nu}^*(\mathbf s) $) for HE settings ($n=1024$ for all)}. These represent the success probabilities of the CC/AI attacks given a compute budget measured in $x$. For an RLWE instance with $\log Q=26$ and a secret with $h=12$, if we run the brute force attack on all secret candidates with up to $x = 5$ cruel bits, the attack would succeed with $30\% $ probability.}
\label{tab:he_probs}
\end{table}

  \begin{table*}[h]
      \centering
      \begin{tabular}{cccc}
      \toprule
          {\bf Paper} & {\bf Attack Type} & {\bf Code link}  & {\bf Language} \\ \midrule
          \cite{Cheon_hybrid_dual} & Dual Hybrid MiTM & \url{https://github.com/swanhong/HybridLWEAttack} & Python and Sage \\ 
          \cite{ducas_sieving} & Sieving & \url{https://github.com/WvanWoerden/G6K-GPU-Tensor} & Python \\ 
          \cite{verde} & ML attack & \url{https://github.com/facebookresearch/verde} & Python \\
            \cite{verde} & uSVP  & 
            \url{https://github.com/facebookresearch/verde} & Python \\ 

            \cite{nolte2024cool} & Cool \& Cruel & \url{https://github.com/facebookresearch/cruel_and_cool} & Python \\ 
            
          \cite{dachman2020lwe} & DBDD & \url{https://github.com/lducas/leaky-LWE-Estimator} & Sage/Python \\
          \cite{fplll} & Lattice reduction & \url{https://github.com/fplll/fplll} & C++/Python \\ 
          \cite{ryan2023fast} & Lattice reduction & \url{https://github.com/keeganryan/flatter} & C++ \\ 
          \cite{rumpsession} & MITM & \url{https://github.com/lducas/leaky-LWE-Estimator/blob/human-LWE/human-LWE/} & Python \\ \bottomrule
        \end{tabular}
    \caption{Available open-source implementations of attacks on Search or Decision LWE as of June 4, 2024.}
    
      \label{tab:github_repos}
  \end{table*}

\subsection{Recoverable Secrets for DH MiTM Attack}
\label{appx:mitm_probs}

%% TODO add back into appendix.
Based on the memory use analysis of \S\ref{subsec:mitm} and Table~\ref{tab:both_mitm_mem}, we can reasonably recover secrets with $h' \le 8$ for all Kyber settings, and $h' \le 8$ for HE settings. One can then easily compute the probability of ``hitting'' secrets with this $h'$ value for an overall secret Hamming weight of $h$, and use this to estimate what Hamming weight secrets are recoverable. These results are recorded in Tables~\ref{tab:he_mitm_probs} and~\ref{tab:kyber_mitm_probs}. 
\begin{table}[h]
\resizebox{0.49\textwidth}{!}{
\centering
\begin{tabular}{lcc|cc|cc}
\toprule
%& \multicolumn{6}{c}{$n = 1024$} \\
 & \multicolumn{2}{c}{$\log q = 26$} & \multicolumn{2}{c}{$\log q = 29$} & \multicolumn{2}{c}{$\log q = 50$} \\
\midrule
%Hamming $h$ & \multirow{2}{*}{8} & \multirow{2}{*}{12} &\multirow{2}{*}{10} &\multirow{2}{*}{12}  & \multirow{2}{*}{17} & \multirow{2}{*}{20}\\ 
% Cruel bits $x$ & & & & & &\\
 & $h=6$ & $h=8$ & $h=7$ & $h=9$ & $h=14$ & $h=16$ \\ 
\midrule
$h'=4$ & 11.4 & 0.5  & 12.9 & 1.5 & 1.0 & 0.2 \\
$h'=6$ & 100.0 & 18.9 & 77.3 & 24.3 & 9.2 & 3.0 \\
$h'=8$ & 100.0 & 100.0 & 100.0 & 85.3 & 39.9 & 19.0 \\
\bottomrule
\end{tabular}
}
\caption{ {\bf Percent chance that secrets with Hamming weight $h$ have $\le h'$ bits in $\zeta$-size MiTM guessing region for HE benchmark settings.} $n=1024$ for all, $\zeta$ given in Table~\ref{tab:both_mitm_mem}. From 10K simulations of secrets.}
\label{tab:he_mitm_probs}
\end{table}

\begin{table}[h]
\resizebox{0.49\textwidth}{!}{
\begin{tabular}{lcc|cc|cc}
\toprule
 & \multicolumn{2}{c}{($k=2$, $\log q = 12$)} & \multicolumn{2}{c}{($k=2$, $\log q = 28$)} & \multicolumn{2}{c}{($k=3$, $\log q = 35$)} \\
\cmidrule{2-7}
 & $h=3$ & $h=4$ & $h=10$ & $h=12$ & $h=12$ & $h=14$ \\ 
\midrule
$h'=4$ & 100.0 & 100.0 & 11.4 & 2.9 & 0.8 & 0.1 \\
$h'=6$ & 100.0 & 100.0 & 53.2 & 23.9 & 11.0 & 2.9 \\
$h'=8$ & 100.0 & 100.0 & 91.9 & 69.2 & 49.3 & 21.3 \\ 
%$h'=10$ & 100.0 & 100.0& 100.0 & 96.8 & 90.8 & 63.7\\ 
\bottomrule
\end{tabular}
}
\caption{ {\bf Percent chance that secrets with Hamming weight $h$ have $\le h'$ bits in $\zeta$-size MiTM guessing region for Kyber benchmark settings.} $n=256$ for all settings, $\zeta$ given in Table~\ref{tab:both_mitm_mem}. From 10K simulations of secrets.}
\label{tab:kyber_mitm_probs}
\end{table}

\subsection{DH MiTM Performance on Small $n$}
\label{sec:mitm_small_n}

Here we present results for smaller $n=128$ LWE setting, with varying $\log_2 q$. For this, we set $\zeta=64$ and $\tau=50$, and run the scaled dual reduction step with $\beta=40$. Table~\ref{tab:mitm_128} presents a summary of results from these experiments: the time required per short vector produced, the estimated bound $B$ for short vectors, and the time required for MiTM attacks on various $h$ secrets. The larger $B$ is as a fraction of $q$, the longer it takes to iterate through all possible secret guesses, because the number of boundary elements to check grows exponentially. Given the difficulty of recovering an $h=10$ secret for a setting with $B / q = 0.08$ and $\zeta=64$, it makes sense that it is difficult to recover MiTM secrets with high $h$ for $\zeta > 500$ and $B/q \approx 0.1$, memory constraints aside. 

\begin{table}[h]

\resizebox{0.49\textwidth}{!}{
\begin{tabular}{llllllll}
\toprule
$\log_2 q$ & 13 & 14 & 15 & 16 & 17 & 18 & 19 \\ \midrule %& 20 \\ \midrule
%\begin{tabular}[c]{@{}l@{}}Reduction \\ time per vector\end{tabular} & 6 mins & 1.8 mins & 3 mins & 7 mins & 3 mins & 8 mins & 4 mins & 4 mins \\ \midrule
$B / q$ & 0.24 & 0.12 & 0.08 & 0.05 & 0.03 & 0.02 & 0.01 \\ \midrule %& 0.01 \\ \midrule
MITM time, $h=5$ &  -  & 5.9 hrs & 25.1s & 5.4s & 0.11s & 0.04s & 0.03s \\  %& 1.3s \\
MiTM time, $h=8$ & - & - & 9 hrs & 11.8 min & 6.0s & 1.0s & 0.4s \\  %& 1.2s \\
MiTM time, $h=10$ &  - & - & 51 hrs & 36 min & 32.3s & 6.5s & 2.7s \\ \bottomrule %& 1.5s \\ \bottomrule
\end{tabular}
}
\caption{{\bf MiTM binary secret recovery times for $n=128$, $\zeta=64$ with varying $\log_2 q$ and $h$.} We include bound $B$/$q$ to demonstrate the relative bound size. Each short vector took $\approx 3.5$ minutes to reduce, using \texttt{flatter} and \texttt{BKZ2.0}, regardless of $\log_2 q$ value. '-' indicates the secret guessing did not finish in $72$ hours, the time limit on our compute cluster.}
\label{tab:mitm_128}
\vspace{-0.3cm}
\end{table}

\subsection{Miscellaneous Tables}
\label{Appendix:misc}

Table~\ref{tab:github_repos} lists open source implementations of LWE attacks available at the time of paper submission. Table~\ref{tab:sieving_mem} estimates memory required for running the GPU implementation of G6K lattice sieiving on dimension $n \ge 128$.

\begin{table}[h]
\centering

\begin{tabular}{cccc} %c}
\toprule
$n$ & \begin{tabular}[c]{@{}c@{}}Max sieving\\ dimension\end{tabular} & \begin{tabular}[c]{@{}c@{}}Max \# of \\ DB vectors\end{tabular} & \begin{tabular}[c]{@{}c@{}}est. DB memory\\ (416 bytes/vector)\end{tabular}
\\ \midrule
128 & 104 & $2^{23.1}$ & 3.6 GB \\
160 & 133 & $2^{29.3}$ & 234 GB \\
256 & 218 & $2^{46.7}$ & 47.8 PB \\
512 & 450 & $2^{94}$ & 1.4e16 PB \\
768 & 682 & $2^{142}$ & 4.6e30 PB\\
1024 & 916 & $2^{191.6}$ & 1.9e45 PB \\
\bottomrule
\end{tabular}
\caption{{\bf Memory estimates for using G6K sieving as the SVP oracle in BKZ}, computed from formulae on pg. 27 of ~\cite{ducas_sieving}. Max sieving dimension is less than $n$ because of the ``dimensions for free'' trick. Database (DB) memory is computed by multiplying estimated $\#$ of database vectors by the reported $416$ bytes/vector storage size on pg. 28 of~\cite{ducas_sieving}.}
\label{tab:sieving_mem}
\end{table}

% % use section* for acknowledgment
% \ifCLASSOPTIONcompsoc
%   % The Computer Society usually uses the plural form
%   \section*{Acknowledgments}
% \else
%   % regular IEEE prefers the singular form
% \fi




\end{document}